\documentclass[onecolumn,showpacs,amsfonts,amsmath,amssymb,floatfix]{revtex4}
\usepackage{url}
\usepackage{bm}		% bold math
\usepackage{graphicx}	% figures
\begin{document}

%\font\tenssf = cmss10
%\input{wasyfont}		% bold math

\title[The interior structure of rotating black holes 3]{The interior structure of rotating black holes 3. Charged black holes}

\author{Andrew J S Hamilton}
\address{JILA and
Dept.\ Astrophysical \& Planetary Sciences,
Box 440, U. Colorado, Boulder, CO 80309, USA}
%\ead{Andrew.Hamilton@colorado.edu}	% iopart
\email{Andrew.Hamilton@colorado.edu}	% revtex
%\homepage{http://casa.colorado.edu/~ajsh/}

\newcommand{\simpropto}{\raisebox{-0.7ex}[1.5ex][0ex]{
		\begin{array}[b]{@{}c@{\;}} \propto \\
		[-1.8ex] \sim \end{array}}}

\newcommand{\dd}{d}
\newcommand{\ddsq}{\dd^2\mkern-1.5mu}
\newcommand{\ddd}{\dd^3\mkern-1.5mu}
\newcommand{\dddd}{\dd^4\mkern-1.5mu}
\newcommand{\DD}{D}
\newcommand{\ee}{e}
\newcommand{\im}{i}
\newcommand{\Ei}{{\rm Ei}}
\newcommand{\perpperp}{\perp\!\!\perp}
\newcommand{\ppartial}{\partial^2\mkern-1mu}
\newcommand{\nn}{\nonumber\\}

\newcommand{\diag}{{\rm diag}}
\newcommand{\jel}{\text{\sl j}}
\newcommand{\Lz}{L}
\newcommand{\Msun}{{\rm M}_\odot}
\newcommand{\uel}{u}
\newcommand{\vel}{v}
\newcommand{\inn}{{\rm in}}
\newcommand{\out}{{\rm ou}}
\newcommand{\sep}{{\rm sep}}

\newcommand{\bg}{\bm{g}}
\newcommand{\bp}{\bm{p}}
\newcommand{\bv}{\bm{v}}
\newcommand{\bx}{\bm{x}}
\newcommand{\bgamma}{\bm{\gamma}}

\newcommand{\Apot}{{\cal A}}
\newcommand{\hatA}{\hat{A}}
\newcommand{\Br}{B}
\newcommand{\betar}{\lambda_r}
\newcommand{\Cx}{C_x}
\newcommand{\Cy}{C_y}
\newcommand{\Dx}{D_x}
\newcommand{\Dy}{D_y}
\newcommand{\Deltax}{\Delta_x}
\newcommand{\Deltaxinf}{\Delta_{x , {\rm inf}}}
\newcommand{\Deltaxsep}{\Delta_{x , {\rm sep}}}
\newcommand{\Deltay}{\Delta_y}
\newcommand{\Er}{E}
\newcommand{\expinf}{\xi}
\newcommand{\Fz}{{\tilde F}}
\newcommand{\starF}{\,{}^\ast\!F}
\newcommand{\starFz}{\,{}^\ast\!\Fz}
\newcommand{\Jx}{J_x}
\newcommand{\Jy}{J_y}
\newcommand{\Mass}{{\cal M}}
\newcommand{\Mbh}{M_\bullet}
\newcommand{\Mdot}{\dot{M}}
\newcommand{\KCarter}{{\cal K}}
\newcommand{\NUT}{{\cal N}}
\newcommand{\px}{p^x}
\newcommand{\Px}{P_x}
\newcommand{\Py}{P_y}
\newcommand{\qbh}{q_\bullet}
\newcommand{\QCarter}{{\cal Q}}
\newcommand{\Qelec}{Q}
\newcommand{\Qelecbh}{\Qelec_\bullet}
\newcommand{\Qmag}{{\cal Q}}
\newcommand{\DeltaQelec}{\Delta \Qelec}
\newcommand{\DeltaQmag}{\Delta \Qmag}
\newcommand{\rhosep}{\rho_{\rm s}}
\newcommand{\rhox}{\rho_x}
\newcommand{\rhoy}{\rho_y}
\newcommand{\Uinf}{U}
\newcommand{\Ur}{{\cal R}}
\newcommand{\Utheta}{\Theta}
\newcommand{\rc}{{\scriptstyle R}}
\newcommand{\tc}{{\scriptstyle T}}
\newcommand{\smallrc}{{\scriptscriptstyle R}}
\newcommand{\smalltc}{{\scriptscriptstyle T}}
\newcommand{\smallzero}{{\scriptscriptstyle 0}}
\newcommand{\Te}{T^{\rm e}}
\newcommand{\Ux}{U_x}
\newcommand{\Uy}{U_y}
\newcommand{\Wx}{W_x}
\newcommand{\Wy}{W_y}
\newcommand{\xin}{x_{\rm in}}
\newcommand{\Zx}{Z_x}
\newcommand{\Zy}{Z_y}
\newcommand{\omegax}{\omega_x}
\newcommand{\omegay}{\omega_y}
\newcommand{\omegaxin}{\omega_{x, {\rm in}}}
\newcommand{\omegayin}{\omega_{y, {\rm in}}}
\newcommand{\omegaxout}{\omega_{x, {\rm out}}}
\newcommand{\omegayout}{\omega_{y, {\rm out}}}

\hyphenpenalty=3000

\begin{abstract}
This paper extends to the case of charged rotating black holes
the conformally stationary, axisymmetric, conformally separable solutions
presented for uncharged rotating black holes in a companion paper.
In the present paper, the collisionless fluid accreted by the
black hole may be charged.
The charge of the black hole is determined self-consistently
by the charge accretion rate.
As in the uncharged case,
hyper-relativistic counter-streaming between ingoing and outgoing streams
drives inflation at (just above) the inner horizon,
followed by collapse.
If both ingoing and outgoing streams are charged,
then conformal separability holds during early inflation,
but fails as inflation develops.
%Two charged streams can exchange energy-momentum
%through the electromagnetic field,
%breaking the $\vel \leftrightarrow - \vel$
%symmetry between ingoing and outgoing streams that seems to be
%fundamental to the existence of conformally separable solutions.
If conformal separability is imposed throughout inflation and collapse,
then only one of the ingoing and outgoing streams can be charged:
the other must be neutral.
Conformal separability prescribes a hierarchy of boundary conditions
on the ingoing and outgoing streams incident on the inner horizon.
The dominant radial boundary conditions require that
the incident ingoing and outgoing number densities be uniform with latitude,
but the charge per particle must vary with latitude such
that the incident charge densities vary
in proportion to the radial electric field.
The sub-dominant angular boundary conditions require specific forms of the
incident number- and charge-weighted angular motions.
If the streams fall freely from outside the horizon,
then the prescribed angular conditions can be achieved by
the charged stream, but not by the neutral stream.
Thus, as in the case of an uncharged black hole,
the neutral stream must be considered to be delivered
ad hoc to just above the inner horizon.
\end{abstract}

	% Classical general relativity

\date{\today}

\maketitle

\section{Introduction}

A companion paper
\cite{Hamilton:2010b},
hereafter Paper~2,
presents conformally stationary,
axisymmetric, conformally separable solutions
for the interior of an uncharged rotating black hole
that undergoes inflation at its inner horizon
and then collapses.
The purpose of this paper is to extend these solutions
to the case of a charged rotating black hole.
%The results of both Paper~2 and the present paper are summarized
%in Paper~1 \cite{Hamilton:2010a}.
A Mathematica notebook containing many details of the calculations
is at \cite{Hamilton:notebook}.

Because of the strength of electromagnetism
and the overall charge neutrality of the Universe,
real astronomical black holes are expected to have little electric charge.
However,
a black hole is likely to build up a residual positive charge
because positively charged protons are more massive than
negatively charged electrons,
so protons are more able to overcome a Coulomb barrier against accretion.
The charge-to-mass ratio of a proton is
$e / m_p \approx 10^{18}$
in Planck units ($c = G = \hbar = 1$).
A black hole might be able to build up a charge-to-mass
of the order of the reciprocal of this ratio
\cite{deDiego:2004ar}.
If so, then trajectories of charged particles
falling into the black hole would be affected by
the black hole's charge notwithstanding its small value.

As shown in Appendix~A of Paper~2,
given the assumptions of conformal time-translation invariance,
axisymmetry, and conformal separability,
the line-element can be taken to be
\begin{equation}
\label{lineelement}
  \dd s^2
  =
  \rho^2
  \left[
  {\dd x^2 \over \Deltax}
  -
  {\Deltax \over \sigma^4}
  \left( \dd t - \omegay \, \dd \phi \right)^2
  +
  {\dd y^2 \over \Deltay}
  +
  {\Deltay \over \sigma^4}
  \left( \dd \phi - \omegax \, \dd t \right)^2
  \right]
  \ ,
\end{equation}
where
$t$ is conformal time,
$\phi$ is the azimuthal coordinate,
$x$ and $y$ are radial and angular coordinates,
$\Deltax$ and $\Deltay$ are radial and angular horizon functions, and
$\sigma \equiv \sqrt{ 1 - \omegax \, \omegay }$.
The conformal factor
$\rho = \rhosep \ee^{\vel t - \expinf}$
is a product of separable (electrovac) $\rhosep$,
time-dependent $\ee^{\vel t}$, and inflationary $\ee^{- \expinf}$ factors.

\section{Collisionless streams}
\label{collisionless}

As in Paper~2 \cite{Hamilton:2010b},
the present paper takes a general freely-falling
collisionless fluid as the source of energy that
ignites and then drives inflation.
In the present paper
collisionless streams are allowed to be electrically charged.

\subsection{Conformal separability conditions}

The tetrad-frame electromagnetic potential $A_k$ is conveniently written
in terms of a set of Hamilton-Jacobi potentials $\Apot_k$
(the following repeats eq.~(24) of Paper~2),
\begin{equation}
\label{Apot}
  A_k
  \equiv
  {1 \over \rho}
  \left\{
  {\Apot_x \over \sqrt{- \Deltax}}
  ,
  {\Apot_t \over \sqrt{- \Deltax}}
  ,
  {\Apot_y \over \sqrt{\Deltay}}
  ,
  {\Apot_\phi \over \sqrt{\Deltay}}
  \right\}
  \ .
\end{equation}
As shown in Appendix~A of Paper~2,
conformal separability requires that
\begin{equation}
\label{fnxy}
  \begin{array}{ccl}
  \omegax
  \ ,
  &
  \Deltax
  &
  \mbox{~are functions of $x$ only}
  \ ,
  \\
  \omegay
  \ ,
  &
  \Deltay
  &
  \mbox{~are functions of $y$ only}
  \ ,
  \end{array}
\end{equation}
and also that
\begin{equation}
\label{fnxyA}
  \begin{array}{ccl}
  \Apot_x
  \ ,
  &
  \Apot_t
  &
  \mbox{~are functions of $x$ only}
  \ ,
  \\
  \Apot_y
  \ ,
  &
  \Apot_\phi
  &
  \mbox{~are functions of $y$ only}
  \ .
  \end{array}
\end{equation}
However,
dimensional analysis shows that
the condition of conformal time-translation symmetry requires that
the potentials $\Apot_k$
must be proportional to the time-dependent factor $\ee^{\vel t}$
of the conformal factor,
\begin{equation}
\label{Apotevt}
  \Apot_k
  \propto
  \ee^{\vel t}
  \ ,
\end{equation}
contradicting conditions~(\ref{fnxyA}).
The dimensional argument is robust;
the proportionality~(\ref{Apotevt}) is correct.
Thus the separability conditions adopted in this paper are,
in place of conditions~(\ref{fnxyA}),
\begin{equation}
\label{fnxyAe}
  \begin{array}{ccl}
  \ee^{- \vel t}
  \Apot_x
  \ ,
  &
  \ee^{- \vel t}
  \Apot_t
  &
  \mbox{~are functions of $x$ only}
  \ ,
  \\
  \ee^{- \vel t}
  \Apot_y
  \ ,
  &
  \ee^{- \vel t}
  \Apot_\phi
  &
  \mbox{~are functions of $y$ only}
  \ .
  \end{array}
\end{equation}

\subsection{Hamilton-Jacobi separation}
\label{HJseparation}

The fact that the conformal separability conditions~(\ref{fnxyA}) fail
and must be replaced by conditions~(\ref{fnxyAe})
implies that the equations of motion of charged particles
in conformally separable spacetimes
are not exactly Hamilton-Jacobi separable.
A similar situation occurred in Paper~2,
where it was found that
the equations of motion of massive particles,
though not exactly Hamilton-Jacobi separable,
are adequately so under the hyper-relativistic conditions of inflation.
This suggests that the Hamilton-Jacobi equations
might still provide an adequate approximation to the equations of motion
of charged particles under the conditions peculiar to inflation.
This subsection shows that the Hamilton-Jacobi equations do in fact
provide an adequate approximation,
but only subject to the special condition~(\ref{Apotzeroparticle}).
Physically,
these conditions require that only one of the ingoing and outgoing streams
can be charged, the other being neutral, \S\ref{onecharge}.

As shown in \S{IV} of Paper~2,
the tetrad-frame momentum $p_k$ of a particle of rest mass $m$ and charge $q$
predicted by the Hamilton-Jacobi equations is
\begin{equation}
\label{pktetrad}
  p_k =
  {1 \over \rho}
  \left\{
  {\Px \over \sqrt{- \Deltax}}
  \, , \ 
  {P_t \over \sqrt{- \Deltax}}
  \, , \ 
  {\Py \over \sqrt{\Deltay}}
  \, , \ 
  {P_\phi \over \sqrt{\Deltay}}
  \right\}
  \ ,
\end{equation}
where the Hamilton-Jacobi parameters $P_t$ and $P_\phi$
are related to the particle's conserved energy
$\pi_t = - E$ and angular momentum $\pi_\phi = L$
and to the potentials $\Apot_t$ and $\Apot_\phi$ by
\begin{equation}
\label{Ptphi}
  P_t
  =
  \pi_t + \pi_\phi \omegax - q \Apot_t
  \ , \quad
  P_\phi
  =
  \pi_\phi + \pi_t \omegay - q \Apot_\phi
  \ ,
\end{equation}
and $\Px$ and $\Py$ are then obtained from
(the following are eqs.~(35) of Paper~2)
\begin{equation}
\label{Pxy}
  \Px
  =
  \sqrt{P_t^2 - \left[ m^2 ( \rho^2 - \rhoy^2 ) + \KCarter \right] \Deltax}
  \ , \quad
  \Py
  =
  \sqrt{- \, P_\phi^2 - ( m^2 \rhoy^2 - \KCarter ) \Deltay}
  \ .
\end{equation}
As it stands,
the tetrad-frame momentum $p_k$
given by equations~(\ref{pktetrad})--(\ref{Pxy})
does not satisfy the Lorentz force law
$\dd p_k / \dd \lambda = q p^l F_{lk}$
to adequate accuracy.

A fix that proves to work under inflationary conditions is,
firstly, to replace
$\Apot_t$ in equation~(\ref{Ptphi}) for $P_t$ by
$( \Apot_t \pm \Apot_x ) / 2$
respectively for ingoing ($+$) and outgoing ($-$) particles,
so that
(given also that $\Apot_\phi = 0$, equation~(\ref{Apotyphizero}))
\begin{equation}
\label{Ptphifix}
  P_t
  \equiv
  \pi_t + \pi_\phi \omegax - q {\Apot_t \pm \Apot_x \over 2}
  \ , \quad
  P_\phi
  =
  \pi_\phi + \pi_t \omegay
  \ ,
\end{equation}
and secondly,
to replace the time-dependent factor $\ee^{\vel t}$ in
$\Apot_t \pm \Apot_x$
by its value as a function of $x$ along the path of the particle
predicted by the Hamilton-Jacobi equations,
\begin{equation}
\label{dtdx}
  {\dd t \over \dd x}
  =
  -
  {1 \over \Px}
  \left( {P_t \over - \Deltax} + {\omegay P_\phi \over \Deltay} \right)
  \ .
\end{equation}
The equation of motion predicted by equations~(\ref{Pxy})--(\ref{dtdx})
is
(the following equation omits the dependency on rest mass $m$,
given previously by eq.~(39) of Paper~2;
the quantities $Z_k$ are defined later, equations~(\ref{Zk}))
\begin{align}
\label{dpkdx}
  {\dd p_k \over \dd \lambda}
  -
  q
  p^l F_{lk}
  &=
  {q \over 2 \rho^3 \sqrt{- \Deltax}}
  \left[
  \vel
  ( \Apot_t \pm \Apot_x ) {P_t \mp \Px \over \Px \Deltax}
  -
  ( \Apot_t \mp \Apot_x )
  \left(
  {\partial \over \partial x}
  \ln \left( {1 \over \sigma^2} {\dd \omegax \over \dd x} \right)
  -
  {P_\phi \over \Px} {\vel \omegay \over \Deltay}
  \right)
  -
  Z_t
  \pm
  Z_x
  \right]
  \left\{
  P_t
  ,
  \Px
  ,
  0
  ,
  0
  \right\}
\nonumber
\\
  &\quad
  + \,
  {q \over 2 \rho^3 \sqrt{\Deltay}}
  ( \Apot_t \mp \Apot_x )
  {1 \over \sigma^2} {\dd \omegay \over \dd y}
  \left\{
  0
  ,
  0
  ,
  P_\phi
  ,
  -
  \Py
  \right\}
  +
  {q \vel \omegay \over \rho^3 \sqrt{- \Deltax \Deltay}}
  {\Apot_x \Px - \Apot_t P_t \over \Px}
  \left\{
  {P_\phi \over \sqrt{\Deltay}}
  \, , \ 
  0
  \, , \ 
  0
  \, , \ 
  {\Px \over \sqrt{- \Deltax}}
  \right\}
  \ .
\end{align}
%\begin{equation}
%  {\dd p_k \over \dd \lambda}
%  -
%  q
%  p^l F_{lk}
%  =
%  {q \vel \over \rho^3 \sqrt{- \Deltax}}
%  \left( \Apot_x \Px - \Apot_t P_t \right)
%  \left\{
%  {1 \over \Px}
%  \left( {P_t \over - \Deltax} + {\omegay P_\phi \over \Deltay} \right)
%  \, , \ 
%  {1 \over - \Deltax}
%  \, , \ 
%  0
%  \, , \ 
%  {\omegay \over \sqrt{- \Deltax \Deltay}}
%  \right\}
%\end{equation}
The right hand side of equation~(\ref{dpkdx}),
which would vanish if
the equations of motion of charged particles were exactly
Hamilton-Jacobi separable,
does not vanish because the Hamilton-Jacobi
approximation~(\ref{Pxy})--(\ref{dtdx}) is not exact.
Appendix~D of Paper~2 gives criteria under which integrals
along the path of a particle may be deemed small,
in the conformally stationary limit.
By these criteria,
all but one of the terms in equation~(\ref{dpkdx})
yields a small result
when integrated over the path of a particle through inflation and collapse.
The discrepant term is the azimuthal $\phi$ component of the last term,
proportional to $\Apot_x \Px - \Apot_t P_t$,
which has $\alpha = -1$ and $\beta = -3$
in the terminology of Appendix~D of Paper~2,
violating (marginally) condition~(D4) during inflation.
Integrated over the path of the particle,
the term would produce a finite difference
between the true azimuthal momentum $p_\phi$
and that predicted by equations~(\ref{Pxy})--(\ref{dtdx}).
The finite difference appears during inflation when $| \Deltax | \ll 1$.
The exception to this conclusion is that the term would vanish
provided that the factor
$\Apot_x \Px - \Apot_t P_t$ multiplying it is zero,
\begin{equation}
\label{ApotP}
  \Apot_x \Px = \Apot_t P_t
  \ .
\end{equation}
But particles are hyper-relativistic,
$P_t = \pm \Px$,
under the conditions $| \Deltax | \ll 1$
where the difference occurs.
Then condition~(\ref{ApotP}) holds provided that
\begin{equation}
\label{Apotzeroparticle}
  \Apot_x = \pm \Apot_t
  \ .
\end{equation}
In other words,
the Hamilton-Jacobi approximation~(\ref{Pxy})--(\ref{dtdx})
works for ingoing particles only if
$\Apot_t = \Apot_x$,
and for outgoing particles only if
$\Apot_t = - \Apot_x$.
Later, \S\ref{onecharge},
it will be concluded that condition~(\ref{Apotzeroparticle})
is equivalent to requiring that only one of the ingoing and outgoing
streams can be charged; the other stream must be neutral.

One might try to go beyond the Hamilton-Jacobi approximation,
but there is no point.
In \S\ref{Tmnelectromagnetic}
it will be found that the non-isotropic diagonal angular component of the
electromagnetic energy-momentum tensor,
which conformal separability requires must vanish,
diverges unless condition~(\ref{Apotzeroparticle}) is true.
Condition~(\ref{Apotzeroparticle}) appears
necessary for the conformally separable solutions
considered in this paper to hold.

\subsection{Electric current}
\label{electric}

Equations governing the density $N$ and number current $n_k$
of a collisionless stream were derived in \S{VII} of Paper~2.
For a single stream of particles of charge $q$ with fixed constants of motion,
the tetrad-frame current $j_k$ is
the particle charge $q$ times the number current $n_k$,
which is itself the number density $N$ times the momentum $p_k$,
\begin{equation}
\label{jkcollisionless}
  j_k
  =
  q n_k
  \ , \quad
  n_k
  =
  N p_k
  \ .
\end{equation}
The Hamilton-Jacobi equations predict that the number density $N$
along a single stream satisfies
\begin{equation}
\label{N}
  N
  \propto
  {\sigma^2 \over \rho^2 P_x P_y}
  \ .
\end{equation}
As discussed in \S\ref{HJseparation},
the equations of motion of charged particles are not exactly
Hamilton-Jacobi separable.
%but are adequately so subject to condition~(\ref{Apotzeroparticle}).
The accuracy of the Hamilton-Jacobi approximation~(\ref{Pxy})--(\ref{dtdx})
can be checked by seeing how closely the covariant divergence
$D^k n_k$ that they predict vanishes.
The result is
(the following equation omits the dependency on rest mass $m$
given previously by eq.~(63) of Paper~2)
\begin{equation}
  D^k n_k
  =
  0
  \ ,
\end{equation}
which happens to vanish identically,
confirming that the Hamilton-Jacobi approximation~(\ref{Pxy})--(\ref{dtdx})
is satisfactorily accurate.

\section{Electromagnetism}
\label{electromagnetism}

Maxwell's equations prove to separate in a manner consistent
with the separation of Einstein's equations
carried out in \S{VIII} of Paper~2 \cite{Hamilton:2010b}.
%Moreover,
%the natural complex structure of the electromagnetic field
%reveals that the conformal factor $\rho$ also has a complex structure,
%equation~(\ref{EiB}).
In this section, the spacetime is taken
to be conformally time-translation symmetric
(not necessarily conformally stationary)
and axisymmetric,
and to satisfy the conformal separability conditions~(\ref{fnxy})
and (\ref{fnxyAe}).

Homogeneous solution of the stationary, separable Einstein equations
leads to the usual electrovac solutions for the
vierbein coefficients $\omegax$ and $\omegay$
of the line-element,
and, as is well-known,
homogeneous solution of the stationary, separable Maxwell equations
leads to the same result
(the following repeats eqs.~(73) of Paper~2):
\begin{equation}
\label{domegadvarpi}
  {\dd \omegax \over \dd x}
  =
  2
  \sqrt{
  \left( f_0 + f_1 \omegax \right)
  \left( g_0 - g_1 \omegax \right)
  }
  \ , \quad
  {\dd \omegay \over \dd y}
  =
  2
  \sqrt{
  \left( f_1 + f_0 \omegay \right)
  \left( g_1 - g_0 \omegay \right)
  }
  \ ,
\end{equation}
where
$f_0$, $f_1$, $g_0$, and $g_1$ are constants set by boundary conditions.
Equations~(\ref{domegadvarpi})
continue to hold through inflation and collapse
in charged as well as neutral black holes.
As found in Paper~2,
inflation occurs generically at an inner horizon
$\Deltax \rightarrow -0$
regardless of the specific choice of the constants
$f_0$, $f_1$, $g_0$, and $g_1$.

\subsection{Electromagnetic field}

The electromagnetic field $F_{mn}$ is a bivector,
and as such has a natural complex structure
\cite{Doran:2003},
with the real part being the electric field,
which changes sign under parity transformation
(a change of sign of all spatial coordinates),
and the imaginary part being the magnetic field,
which is unchanged by a parity transformation.
The complex structure is manifest in a complexified electromagnetic field
$\Fz_{mn}$ defined by
\begin{equation}
\label{complexifiedemfield}
  \Fz_{kl}
  \equiv
  \frac{1}{2}
  \left(
  F_{kl}
  +
  \starF_{kl}
  \right)
  \ ,
\end{equation}
where $\starF_{kl}$ denotes the Hodge dual of $F_{kl}$,
\begin{equation}
\label{dualemfield}
  \starF_{kl}
  \equiv
  \frac{\im}{2} \,
  \varepsilon_{kl}{}^{mn} \,
  F_{mn}
  \ ,
\end{equation}
with
$\varepsilon_{klmn}$
the totally antisymmetric tensor,
normalized here to
$\varepsilon^{klmn} = [klmn]$
in an orthonormal tetrad frame.
The complexified electromagnetic field tensor is self-dual,
$\starFz_{kl} = \Fz_{kl}$.
Given conformal time-translation symmetry
(not necessarily conformally stationarity),
axisymmetry,
and conformal separability,
the tetrad-frame complexified electromagnatic field tensor $\Fz_{mn}$ is
\begin{subequations}
\label{Fz}
\begin{align}
\label{Fzxt}
  \Fz_{xt}
  \equiv
  \frac{1}{2}
  \left(
  F_{xt}
  +
  \im
  F_{\phi y}
  \right)
  &=
  {1 \over 2 \rho^2}
  \left[
  -
  \left(
  {\partial \Apot_t \over \partial x}
  +
  {\omegay \Apot_t - \Apot_\phi \over \sigma^2}
  {\dd \omegax \over \dd x}
  \right)
  +
  \im
  \left(
  {\partial \Apot_\phi \over \partial y}
  +
  {\omegax \Apot_\phi - \Apot_t \over \sigma^2}
  {\dd \omegay \over \dd y}
  \right)
  -
  \vel
  \left(
  {\Apot_x \over \Deltax}
  +
  \im
  {\omegay \Apot_y \over \Deltay}
  \right)
  \right]
%\\
%  &=
%  -
%  {1 \over 2 \rho^2}
%  \left\{
%  \Apot_t
%  \left[
%  {\partial \over \partial x}
%  \ln
%  \left(
%  {1 \over \sigma^2}
%  {\dd \omegax \over \dd x}
%  \right)
%  +
%  {\im \over \sigma^2}
%  {\dd \omegay \over \dd y}
%  \right]
%  +
%  \Apot_\phi
%  \left[
%  {\partial \over \partial y}
%  \ln
%  \left(
%  {1 \over \sigma^2}
%  {\dd \omegay \over \dd y}
%  \right)
%  -
%  {\im \over \sigma^2}
%  {\dd \omegax \over \dd x}
%  \right]
%  +
%  Z_t
%  -
%  \im
%  Z_\phi
%  \right\}
%\\
%  &=
%  {1 \over 2 \rho^2}
%  \left[
%  {\rhosep^2 ( \Qelec + \im \Qmag ) \over
%  ( \rhox
%  - \im
%  \rhoy
%  )^2}
%  +
%  Z_t
%  -
%  \im
%  Z_\phi
%  \right]
  \ ,
\\
\label{Fzxy}
  \Fz_{xy}
  \equiv
  \frac{1}{2}
  \left(
  F_{xy}
  +
  \im
  F_{t\phi}
  \right)
  &=
  \im
  {\vel
  ( \omegay \Apot_t - \Apot_\phi )
  \over
  2 \rho^2
  \sqrt{- \Deltax \Deltay}}
  \ ,
\\
\label{Fzxphi}
  \Fz_{x\phi}
  \equiv
  \frac{1}{2}
  \left(
  F_{x\phi}
  +
  \im
  F_{yt}
  \right)
  &=
  {\vel
  (
  \omegay \Apot_x
  + \im
  \Apot_y
  )
  \over
  2 \rho^2
  \sqrt{- \Deltax \Deltay}}
  \ .
\end{align}
\end{subequations}
If the spacetime were strictly stationary, $\vel \equiv 0$,
then the only non-vanishing component of the complexified
electromagnetic field would be the radial component $\Fz_{xt}$.
In inflationary spacetimes, however,
the radial horizon function $\Deltax$ goes to zero at the inner horizon,
and the angular components
$\Fz_{xy}$ and $\Fz_{x\phi}$
of the electromagnetic field can grow large at the inner horizon,
however small the accretion rate $\vel$ may be.

The only electromagnetic gauge freedom
that respects conformal time-translation symmetry
and the conformal separability conditions~(\ref{fnxyAe}) is
$A_k \rightarrow A_k + \lambda \, \partial_k \ee^{\vel t}$
for some constant $\lambda$,
which transforms
\begin{equation}
\label{Agauge}
  \Apot_t
  \rightarrow
  \Apot_t
  +
  \lambda \vel
  \ee^{\vel t}
  \ , \quad
  \Apot_\phi
  \rightarrow
  \Apot_\phi
  +
  \lambda \vel \omegay
  \ee^{\vel t}
  \ .
\end{equation}

Define the enclosed electric charge $\Qelec$ within radius $x$,
and the enclosed magnetic charge $\Qmag$ above latitude $y$, by
\begin{subequations}
\label{Qelecmag}
\begin{align}
%\fl
\label{Qelec}
  \Qelec
  &\equiv
  -
  2 ( f_0 g_1 + f_1 g_0 )
  {\Apot_t \over {\dd \omegax / \dd x}}
  \ ,
\\
\label{Qmag}
  \Qmag
  &\equiv
  -
  2 ( f_0 g_1 + f_1 g_0 )
  {\Apot_\phi \over {\dd \omegay / \dd y}}
  \ .
\end{align}
\end{subequations}
%The combination
%$f_0 g_1 + f_1 g_0$
%of constants is positive, without loss of generality.
Further, define the quantities
$Z_k$ by
\begin{subequations}
\label{Zk}
\begin{align}
\label{Zxt}
  Z_x
  \equiv
  {\dd \omegax \over \dd x}
  {\partial \over \partial x}
  \left(
  {\Apot_x \over \dd \omegax / \dd x}
  \right)
  +
  {\vel \Apot_t \over \Deltax}
  \ &, \quad
  Z_t
  \equiv
  {\dd \omegax \over \dd x}
  {\partial \over \partial x}
  \left(
  {\Apot_t \over \dd \omegax / \dd x}
  \right)
  +
  {\vel \Apot_x \over \Deltax}
  \ ,
\\
\label{Zyphi}
  Z_y
  \equiv
  {\dd \omegay \over \dd y}
  {\partial \over \partial y}
  \left(
  {\Apot_y \over \dd \omegay / \dd y}
  \right)
  +
  {\vel \omegay \Apot_\phi \over \Deltay}
  \ &, \quad
  Z_\phi
  \equiv
  {\dd \omegay \over \dd y}
  {\partial \over \partial y}
  \left(
  {\Apot_\phi \over \dd \omegay / \dd y}
  \right)
  -
  {\vel \omegay \Apot_y \over \Deltay}
  \ .
\end{align}
\end{subequations}
The conventional radial electric and magnetic fields $\Er$ and $\Br$
constitute the real and imaginary parts of (twice) the
radial electromagnetic field.
The radial electromagnetic field
$\Fz_{xt}$
can be written in terms of the enclosed electric and magnetic charges
$\Qelec$ and $\Qmag$
and the $Z_k$ as
\begin{equation}
\label{EiBelec}
  2 \Fz_{xt}
  =
  \Er + \im \Br
  =
  {1 \over \rho^2}
  \left[
  ( \Qelec + \im \Qmag )
  {\rhox + \im \rhoy
  \over
  \rhox - \im \rhoy}
  -
  Z_t
  +
  \im
  Z_\phi
  \right]
  \ ,
\end{equation}
where
$\rhox$ and $\rhoy$ are the $x$ and $y$ components
of the separable conformal factor
$\rhosep = \sqrt{ \rhox^2 + \rhoy^2}$,
equation~(72) of Paper~2.

\subsection{Maxwell's equations}

Maxwell's equations are embodied in the complex equation
\begin{equation}
\label{Maxwell}
  \DD^m \Fz_{mn}
  =
  2\pi j_n
  \ ,
\end{equation}
whose real (electric) and imaginary (magnetic) parts constitute respectively
the source and source-free Maxwell's equations.
Given conformal time-translation invariance
and the conformal separability conditions~(\ref{fnxy}) and (\ref{fnxyAe}),
the source-free Maxwell's equations are satisfied identically
with vanishing magnetic current.
Since the magnetic current,
and in particular its time component the magnetic charge,
necessarily vanishes,
the solutions preclude the accretion of any magnetic charge.
Although strictly stationary solutions admit a black hole with magnetic charge,
conformally stationary solutions do not.

In terms of the $Z_k$ defined by equations~(\ref{Zk}),
and given equations~(\ref{domegadvarpi})
for $\dd \omegax / \dd x$ and $\dd \omegay / \dd y$,
the sourced Maxwell's equations can be written
\begin{subequations}
\label{MaxwellZ}
\begin{align}
\label{MaxwellZxt}
  j_x \pm j_t
  &=
  {1 \over 4\pi \rho^3}
  \left\{
  {\vel \over \sqrt{- \Deltax}}
  \left[
  ( \Apot_x \pm \Apot_t )
  \left(
  \pm \,
  {\partial \over \partial x}
  \ln \left( {1 \over \sigma^2} {\dd \omegax \over \dd x} \right)
  -
  {\vel \omegay^2 \over \Deltay}
  \right)
  \pm
  \Apot_y
  {\partial \over \partial y}
  \ln \left( {1 \over \sigma^2} {\dd \omegay \over \dd y} \right)
  -
  \Apot_\phi
  {1 \over \sigma^2} {\dd \omegax \over \dd x}
  \right]
  \right.
\nonumber
\\
  &\quad
  \left.
  + \,
  {\vel \over \sqrt{- \Deltax}}
  ( Z_t \pm Z_y )
  \pm
  \sqrt{- \Deltax}
  \left[
  {\partial Z_t \over \partial x}
  +
  Z_t
  {\partial \over \partial x}
  \ln \left( {1 \over \sigma^2} {\dd \omegax \over \dd x} \right)
  -
  Z_\phi
  {1 \over \sigma^2} {\dd \omegay \over \dd y}
  \right]
  \right\}
  \ ,
\\
\label{MaxwellZyphi}
  j_y \pm \im j_\phi
  &=
  {1 \over 4\pi \rho^3}
  \left\{
  {\vel \omegay \over \sqrt{\Deltay}}
  \left[
  ( \Apot_y \pm \im \Apot_\phi )
  \left(
  \mp \, \im
  {\partial \over \partial y}
  \ln
  \left(
  {1 \over \sigma^2} {\dd \omegay \over \dd y}
  \right)
  +
  {\vel \over \omegay \Deltax}
  \right)
  \mp
  \im
  \Apot_x
  {\partial \over \partial x}
  \ln
  \left(
  {1 \over \sigma^2} {\dd \omegax \over \dd x}
  \right)
  -
  \Apot_t
  {1 \over \sigma^2} {\dd \omegay \over \dd y}
  \right]
  \right.
\nonumber
\\
  &\quad
  \left.
  + \,
  {\vel \omegay \over \sqrt{\Deltay}}
  ( Z_\phi \mp \im Z_x )
  \mp
  \im
  \sqrt{\Deltay}
  \left[
  {\partial Z_\phi \over \partial y}
  +
  Z_\phi
  {\partial \over \partial y}
  \ln
  \left(
  {1 \over \sigma^2} {\dd \omegay \over \dd y}
  \right)
  -
  Z_t
  {1 \over \sigma^2} {\dd \omegax \over \dd x}
  \right]
  \right\}
  \ .
\end{align}
\end{subequations}
These equations yield both electrovac and inflationary solutions.

Electrovac solutions correspond to the case of strict stationarity,
$\vel = 0$.
In strictly stationary spacetimes,
$\Apot_x$ and $\Apot_y$ can be set to zero by a gauge transformation,
as is evident from the fact that for $\vel = 0$
the electromagnetic field, equations~(\ref{Fz}),
is independent of
$\Apot_x$ and $\Apot_y$.
This gauge freedom is available in strictly stationary
but not conformally stationary spacetimes.
The homogeneous solutions of Maxwell's equations~(\ref{MaxwellZ})
are those with $Z_t = Z_\phi = 0$,
which given the definitions~(\ref{Zk}) of $Z_k$,
correspond to spacetimes with constant electric and magnetic
charges $\Qelec$ and $\Qmag$, equations~(\ref{Qelecmag}).

Inflationary solutions of Maxwell's equations~(\ref{MaxwellZ})
have small but non-zero accretion rate $\vel$.
During inflation,
the angular currents
$j_y$ and $j_\phi$ 
available from a collisionless source are small.
The term proportional to $\Apot_y + \im \Apot_\phi$
in equation~(\ref{MaxwellZyphi})
involves a factor of $\vel / \Deltax$
which diverges at the inner horizon $\Deltax \rightarrow -0$
however small the accretion accretion rate $\vel$ might be.
The only way that this term can remain small is that
\begin{equation}
\label{Apotyphizero}
  \Apot_y = \Apot_\phi = 0
  \ ,
\end{equation}
which is equivalent to requiring that the magnetic charge be identically zero
(an apparent exception to this argument is that if
$\Apot_\phi$
is chosen to be a constant times $\omegay$,
then the divergent term can cancel against a corresponding term
proportional to $\Apot_t$ in $Z_x$;
but that simply reflects the gauge freedom~(\ref{Agauge}) in $\Apot_\phi$).
%so $\Apot_\phi$ can be set to zero without loss of generality).
That magnetic charge must vanish
accords with the conclusion at the beginning of this subsection,
that conformal time-translation invariance and conformal separability
force the magnetic current to vanish identically,
so the black hole cannot accrete magnetic charge,
so its cumulative magnetic charge must be zero.

Given the vanishing of $\Apot_y$ and $\Apot_\phi$,
it follows that $Z_y$ and $Z_\phi$ vanish identically,
and Maxwell's equations~(\ref{MaxwellZ}) reduce to
\begin{subequations}
\label{MaxwellZnomag}
\begin{align}
\label{MaxwellZnomagxt}
  4\pi \rho^2 ( j_x \pm j_t )
  &=
  {1 \over \rho}
  \left\{
  {\vel \over \sqrt{- \Deltax}}
  ( \Apot_x \pm \Apot_t )
  \left[
  \pm \,
  {\partial \over \partial x}
  \ln \left( {1 \over \sigma^2} {\dd \omegax \over \dd x} \right)
  -
  {\vel \omegay^2 \over \Deltay}
  \right]
  \pm
  \sqrt{- \Deltax}
  \left[
  {\partial \over \partial x}
  \mp
  {\vel \over \Deltax}
  +
  {\partial \over \partial x}
  \ln \left( {1 \over \sigma^2} {\dd \omegax \over \dd x} \right)
  \right]
  Z_t
  \right\}
  \ ,
\\
\label{MaxwellZnomagyphi}
  4\pi \rho^2 ( j_y \pm \im j_\phi )
  &=
  {1 \over \rho}
  \left\{
  {\vel \omegay \over \sqrt{\Deltay}}
  \left[
  \mp
  \im
  \Apot_x
  {\partial \over \partial x}
  \ln
  \left(
  {1 \over \sigma^2} {\dd \omegax \over \dd x}
  \right)
  -
  \Apot_t
  {1 \over \sigma^2} {\dd \omegay \over \dd y}
  \right]
  \mp
  \im
  \left(
  Z_x
  {\vel \omegay
  \over \sqrt{\Deltay}}
  -
  Z_t
  \sqrt{\Deltay}
  {1 \over \sigma^2} {\dd \omegax \over \dd x}
  \right)
  \right\}
  \ .
\end{align}
\end{subequations}

\subsection{Electromagnetic energy-momentum}
\label{Tmnelectromagnetic}

In terms of the complexified electromagnetic field
$\Fz_{mn}$,
the tetrad-frame electromagnetic energy-momentum tensor
$\Te_{kl}$
satisfies
\begin{equation}
\label{Tem}
  4\pi \Te_{kl}
  =
  \eta^{mn}
  \left(
  \Fz_{km} \Fz^\ast_{ln}
  +
  \Fz_{kn} \Fz^\ast_{lm}
  \right)
  \ ,
\end{equation}
in which $^\ast$ denotes the complex conjugate
(not the Hodge dual).
Given that the angular electromagnetic potentials $\Apot_y$ and $\Apot_\phi$
vanish,
the tetrad-frame electromagnetic energy-momentum tensor
$\Te_{kl}$
is
\begin{subequations}
\label{Tmnem}
\begin{align}
\label{Txxmttem}
  \Te_{xx} - \Te_{tt}
  =
  \Te_{yy} + \Te_{\phi\phi}
  &=
  {\Er^2 + \Br^2 \over 4\pi}
  \ ,
\\
\label{Txtem}
  {\Te_{xx} + \Te_{tt} \over 2} \pm \Te_{xt}
  &=
  -
  {\vel^2 \omegay^2 ( \Apot_x \pm \Apot_t )^2
  \over 8\pi \rho^4 \Deltax \Deltay}
  \ ,
\\
\label{Txyem}
  \Te_{xy} \pm \Te_{ty}
  &=
  -
  {\vel \omegay \over 4\pi \rho^2 \sqrt{-\Deltax \Deltay}}
  B ( \Apot_x \pm \Apot_t )
  \ ,
\\
\label{Txphiem}
  \Te_{x\phi} \pm \Te_{t\phi}
  &=
  \mp
  {\vel \omegay \over 4\pi \rho^2 \sqrt{-\Deltax \Deltay}}
  E ( \Apot_x \pm \Apot_t )
  \ ,
\\
\label{Typhiem}
  {\Te_{yy} - \Te_{\phi\phi} \over 2} \pm \im \Te_{y\phi}
  &=
  -
  {\vel^2 \omegay^2 \over 8\pi\rho^4 \Deltax \Deltay}
  ( \Apot_x + \Apot_t ) ( \Apot_x - \Apot_t )
  \ ,
\end{align}
\end{subequations}
where $\Er$ and $\Br$
are the radial electric and magnetic fields from equation~(\ref{EiBelec}).

As discussed in \S{VIII\,I} of Paper~2,
the condition of conformal separability requires that
the $2 \times 2$ angular submatrix of the energy-momentum tensor
must be isotropic
(proportional to the unit $2 \times 2$ unit matrix),
since the non-isotropic angular Einstein components depend only on angle $y$
(modulo an overall conformal factor),
and being initially negligible in the conformally stationary limit,
must remain so at any radius $x$.
The isotropy of the angular energy-momentum
requires that the components given by equation~(\ref{Typhiem}) must
be negligible.
On the other hand
the right hand side of expression~(\ref{Typhiem})
is proportional to $1/\Deltax$,
which diverges at the inner horizon $\Deltax \rightarrow -0$.
The only way out is
that one of the remaining factors in the expression must vanish,
which requires that
\begin{equation}
\label{Apotzero}
  \Apot_x = \pm \Apot_t
  \ .
\end{equation}
Condition~(\ref{Apotzero}) is the same as that~(\ref{Apotzeroparticle})
found previously in order that the equations of motion of charged particles
be Hamilton-Jacobi separable to adequate accuracy.

\section{Inflationary solutions}

This section presents inflationary solutions to the combined
Maxwell and Einstein equations.
Subsection~\ref{evolutionA}
derives the evolution of the electric potentials $\Apot_x$ and $\Apot_t$
from the vanishing of $Z_x$ and $Z_t$,
which Maxwell's equations sourced by a collisionless current
require to be small.
The case of small but finite $Z_k$ is deferred to Appendix~\ref{Zsec}.
Subsection~\ref{collisionlesscurrentsource}
shows that, with these potentials,
Maxwell's equations are satisfied by a sum of currents
from ingoing and outgoing collisionless streams.
Subsection~\ref{onecharge}
concludes that only one or other of the ingoing and outgoing streams
can be charged.
Subsection~\ref{Einstein8}
reviews 8 of the Einstein components,
and \S\ref{collisionlessenergymomentumsource}
shows that the energy-momentum required by these 8 Einsteins,
after subtraction of the electromagnetic energy-momentum,
is satisfied by the
energy-momentum of ingoing and outgoing collisionless streams.
The remaining 2 Einstein components were shown in \S{VIII\,D,E} of Paper~2
to govern the evolution of the inflationary exponent $\expinf$ and
horizon function $\Deltax$.
Subsection~\ref{Einstein2},
following along the lines of \S{VIII\,J} of Paper~2,
shows how the sub-dominant electromagnetic source
for these 2 Einstein components can be taken into account by solving
the Einstein equations to next higher order.

\subsection{Evolution of the electromagnetic potential and enclosed charge}
\label{evolutionA}

The definitions~(\ref{Zk}) of $Z_k$ provide
evolutionary equations for the electric potentials $\Apot_x$ and $\Apot_t$:
\begin{equation}
\label{dApot}
  \left(
  {\partial \over \partial x}
  \pm
  {\vel \over \Deltax}
  \right)
  {\Apot_x \pm \Apot_t \over \dd \omegax / \dd x}
  =
  {Z_x \pm Z_t \over \dd \omegax / \dd x}
  \ .
\end{equation}
Since the angular components of the collisionless current must be small,
Maxwell's equations~(\ref{MaxwellZnomagyphi})
require that $Z_x$ and $Z_t$ be small.
The dominant driving term
in equation~(\ref{dApot})
is then the one proportional to $\vel / \Deltax$,
which diverges at the inner horizon $\Deltax \rightarrow -0$
however small the accretion rate $\vel$ may be.
By comparison, the effect of a small but finite $Z_x$ and $Z_t$
is essentially negligible.
In Appendix~\ref{Zsec}
the effect of small but finite $Z_k$ will be considered,
but for the remainder of this paper
$Z_x$ and $Z_t$
will be taken to vanish:
\begin{equation}
\label{Zzero}
  Z_x = Z_t = 0
  \ .
\end{equation}

Thus the solutions of interest are
the homogeneous solutions of equations~(\ref{dApot}),
those with vanishing right hand side.
Equations~(\ref{dApot})
depend on the horizon function $\Deltax$,
whose behaviour as a function of radius $x$ was solved in Paper~2.
Equation~(93a) of Paper~2 gives
\begin{equation}
\label{dxD}
  {\dd x \over \Deltax}
  =
  - {\dd \Ux \over 2 ( \Ux^2 - \vel^2 )}
  =
  {1 \over 4 \vel} \dd \ln \left( {\Ux + \vel \over \Ux - \vel} \right)
  \ .
\end{equation}
%\begin{equation}
%  \Deltax
%  {\partial \over \partial x}
%  =
%  - 2 ( \Ux^2 - \vel^2 )
%  {\partial \over \partial \Ux}
%  \ .
%\end{equation}
Consequently the homogeneous solutions of equations~(\ref{dApot}) are
\begin{equation}
\label{ApotQ}
  {\Apot_x \pm \Apot_t \over \dd \omegax / \dd x}
  =
  \mp
  {\Qelec^\pm \over 4 ( f_0 g_1 + f_1 g_0 )}
  \ ,
\end{equation}
where $Q^\pm$ are ingoing ($+$) and outgoing ($-$)
enclosed electric charges satisfying
\begin{equation}
\label{Qpm}
  \Qelec^\pm
  =
  \Qelecbh^\pm
  \ee^{\vel t}
  \left[
  ( \Ux - \vel ) ( \uel + \vel )
  \over
  ( \Ux + \vel ) ( \uel - \vel )
  \right]^{\pm 1/4}
  \ ,
\end{equation}
with $\Qelecbh^\pm$ constants of integration.
The total enclosed electric charge $\Qelec$,
which is related to the potential $\Apot_t$ by
equation~(\ref{Qelec}),
is a sum of the ingoing and outgoing enclosed charges,
\begin{equation}
  \Qelec
  =
  \Qelec^+ + \Qelec^-
  \ .
\end{equation}
The constants
$\Qelecbh^\pm$
physically represent the cumulative ingoing and outgoing electric charge
accreted by the black hole up to $t = 0$.
The total electric charge $\Qelecbh$
of the black hole at $t = 0$,
as seen by an observer well outside the horizon,
is a sum of the ingoing and outgoing cumulative charges,
\begin{equation}
  \Qelecbh
  =
  \Qelecbh^+
  +
  \Qelecbh^-
  \ .
\end{equation}

The relation between time $t$ and radius $x$ along the path
of a particle is given by equation~(\ref{dtdx}).
In the hyper-relativistic conditions $P_t = \pm \Px$
characteristic of inflation and collapse when $| \Deltax | \ll 1$,
equation~(\ref{dtdx}) simplifies to $\dd t / \dd x = \pm 1 / \Deltax$,
which given equation~(\ref{dxD}) integrates to
\begin{equation}
\label{evtx}
  \ee^{\vel t}
  =
  \left[
  ( \Ux + \vel ) ( \uel - \vel )
  \over
  ( \Ux - \vel ) ( \uel + \vel )
  \right]^{\pm 1/4}
  \ .
\end{equation}
The relation~(\ref{evtx}) continues to hold even at the end of collapse,
when $| \Deltax |$ ceases to be small,
and $\Ux$ is growing exponentially huge,
and the time coordinate $t$ is frozen.

The solutions~(\ref{Qpm}) for the enclosed electric charges $\Qelec^\pm$
have the salient feature that,
in view of equation~(\ref{evtx}),
the ingoing charge $\Qelec^+$
is constant along the path of an ingoing particle,
while
the outgoing charge $\Qelec^-$
is constant along the path of an outgoing particle.
As found in \S{VIII\,E} of Paper~2,
in the conformally stationary limit
the coordinates $x$ and $y$ along the path of a freely-falling stream
remain frozen throughout inflation and collapse,
so the factor $\dd \omegax / \dd x$
in the relation~(\ref{ApotQ}) is constant.
Consequently the potentials $\Apot_x \pm \Apot_t$
are also constant along the paths of respectively
ingoing and outgoing particles.
Among other things, this implies that the Hamilton-Jacobi parameter
$P_t$ defined by equation~(\ref{Ptphifix}) is constant
along the path of an ingoing or outgoing particle.
Likewise the Hamilton-Jacobi parameter $\Px$ defined by equation~(\ref{Pxy})
is constant along the path of an ingoing or outgoing particle
during inflation and collapse, as long as $| \Deltax | \ll 1$.

%In the hyper-relativistic conditions characteristic of inflation,
%ingoing and outgoing particles have $p_t = \pm p_x$,
%and thus ingoing and outgoing currents have $j_t = \pm j_x$,
%at least until angular motions become important at the end of collapse.
Maxwell's equations~(\ref{MaxwellZnomag})
show that the potentials $\Apot_x \pm \Apot_t$
are sourced respectively by ingoing and outgoing currents,
and may thus be called ingoing and outgoing potentials.
To bring out the dependence on the ingoing and outgoing potentials
$\Apot_x \pm \Apot_t$,
it is helpful to reexpress
Maxwell's equations~(\ref{MaxwellZnomag})
in terms of a sum of ingoing and outgoing currents
\begin{equation}
\label{jk}
  j_k
  =
  j^+_k
  +
  j^-_k
  \ ,
\end{equation}
where for vanishing $Z_k$, equation~(\ref{Zzero}),
%\begin{subequations}
%\label{MaxwellZzero}
%\begin{align}
%\label{MaxwellZzeroxt}
%  j_x \pm j_t
%  &=
%  {1 \over 4\pi \rho^3}
%  {\vel \over \sqrt{- \Deltax}}
%  ( \Apot_x \pm \Apot_t )
%  \left[
%  \pm \,
%  {\partial \over \partial x}
%  \ln \left( {1 \over \sigma^2} {\dd \omegax \over \dd x} \right)
%  -
%  {\vel \omegay^2 \over \Deltay}
%  \right]
%  \ ,
%\\
%\label{MaxwellZzeroyphi}
%  j_y \pm \im j_\phi
%  &=
%  {1 \over 4\pi \rho^3}
%  {\vel \omegay \over \sqrt{\Deltay}}
%  \left[
%  \mp \,
%  \im
%  \Apot_x
%  {\partial \over \partial x}
%  \ln
%  \left(
%  {1 \over \sigma^2} {\dd \omegax \over \dd x}
%  \right)
%  -
%  \Apot_t
%  {1 \over \sigma^2} {\dd \omegay \over \dd y}
%  \right]
%  \ .
%\end{align}
%\end{subequations}
\begin{subequations}
\label{MaxwellZzero}
\begin{align}
\label{MaxwellZzeroxt}
  j^\pm_x
  =
  \pm
  j^\pm_t
  &=
  {1 \over 8\pi \rho^3}
  {\vel \over \sqrt{- \Deltax}}
  ( \Apot_x \pm \Apot_t )
  \left[
  \pm \,
  {\partial \over \partial x}
  \ln \left( {1 \over \sigma^2} {\dd \omegax \over \dd x} \right)
  -
  {\vel \omegay^2 \over \Deltay}
  \right]
  \ ,
\\
\label{MaxwellZzeroy}
  j^\pm_y
  &\equiv
  \mp
  {1 \over 8\pi \rho^3}
  {\vel \omegay \over \sqrt{\Deltay}}
  ( \Apot_x \pm \Apot_t )
  {1 \over \sigma^2} {\dd \omegay \over \dd y}
  \ ,
\\
\label{MaxwellZzerophi}
  j^\pm_\phi
  &\equiv
  -
  {1 \over 8\pi \rho^3}
  {\vel \omegay \over \sqrt{\Deltay}}
  ( \Apot_x \pm \Apot_t )
  {\partial \over \partial x}
  \ln
  \left(
  {1 \over \sigma^2} {\dd \omegax \over \dd x}
  \right)
  \ .
\end{align}
\end{subequations}
Should not the term proportional to $\vel$ inside square brackets
on the right hand side of equation~(\ref{MaxwellZzeroxt})
be neglected compared to the dominant first term,
in the conformally stationary limit $\vel \rightarrow 0$?
No.
The term is needed to ensure that the equations hold not
only to leading radial order but also to sub-dominant angular order.

\subsection{Collisionless source of electric current}
\label{collisionlesscurrentsource}

Maxwell's equations~(\ref{jk})--(\ref{MaxwellZzero})
can be satisfied by currents from
a sum of ingoing and outgoing collisionless streams,
\begin{equation}
\label{jpm}
  j^\pm_k
  =
  q^\pm N^\pm p^\pm_k
  \ ,
\end{equation}
%but only if one of the streams is charged and the other is neutral.
%If the charged stream is ingoing $(+)$,
%then $\Apot_x + \Apot_t$ is constant along the path of the stream,
%while if it is outgoing,
%then $\Apot_x - \Apot_t$ is constant along the path of the stream.
%The charged stream has charge density
with charge densities
\begin{equation}
\label{qNpm}
  q^\pm N^\pm
  =
  {\vel \Qelec^\pm \over 32\pi \rho^2 ( f_0 g_1 + f_1 g_0 )}
  {\dd \omegax \over \dd x}
  \left[
  {\partial \over \partial x}
  \ln \left( {1 \over \sigma^2} {\dd \omegax \over \dd x} \right)
  \mp
  {\vel \omegay^2 \over \Deltay}
  \right]
  \ ,
\end{equation}
and hyper-relativistic tetrad-frame momenta
\begin{equation}
\label{qppm}
  p^\pm_k
  =
  {1 \over \rho}
  \left\{
  -
  {1 \over \sqrt{- \Deltax}}
  \, , \ 
  \mp
  {1 \over \sqrt{- \Deltax}}
  \, , \ 
  {1 \over \sqrt{\Deltay}}
  {
  \displaystyle
  {1 \over \sigma^2} {\dd \omegay \over \dd y}
  \over
  \displaystyle
  {\partial \over \partial x}
  \ln \left( {1 \over \sigma^2} {\dd \omegax \over \dd x} \right)
  \mp
  {\vel \omegay^2 \over \Deltay}}
  \, , \ 
  \pm
  {1 \over \sqrt{\Deltay}}
  {
  \displaystyle
  {\partial \over \partial x}
  \ln
  \left(
  {1 \over \sigma^2} {\dd \omegax \over \dd x}
  \right)
  \over
  \displaystyle
  {\partial \over \partial x}
  \ln \left( {1 \over \sigma^2} {\dd \omegax \over \dd x} \right)
  \mp
  {\vel \omegay^2 \over \Deltay}
  }
  \right\}
\ .
\end{equation}
The densities~(\ref{qNpm}) and momenta~(\ref{qppm})
are defined up to arbitrary normalization factors
such that their product is constant.
The densities~(\ref{qNpm}) and momenta~(\ref{qppm})
conform with the behaviour of collisionless streams
during inflation and collapse as long as
\begin{equation}
\label{Dsmall}
  | \Deltax | \ll 1
  \ ,
\end{equation}
which is to say
before the angular motions of the streams become important.
The Hamilton-Jacobi result~(\ref{N})
requires that the density along a stream evolve as
\begin{equation}
\label{Npmprop}
  N \propto {1 \over \rho^2}
  \ ,
\end{equation}
since the Hamilton-Jacobi parameters $\Px$ and $\Py$ are constant
and $\sigma$ is frozen at its inner horizon value.
The densities from equation~(\ref{qNpm}) indeed satisfy
the proportionality~(\ref{Npmprop}),
since all other factors in the equation are constant along the
path of the stream
(including $\Qelec^\pm$, as shown in the previous subsection~\ref{evolutionA}).
Similarly the tetrad-frame momentum $p^\pm_k$, equation~(\ref{qppm}),
accords with the Hamilton-Jacobi form~(\ref{pktetrad}),
with constant Hamilton-Jacobi parameters $P_k$
along the path of the stream.

\subsection{Only one stream can be charged}
\label{onecharge}

In \S\ref{HJseparation},
the condition
$\Apot_t = \pm \Apot_x$,
equation~(\ref{Apotzeroparticle}),
emerged from requiring that the motions of charged particle
be adequately described by the Hamilton-Jacobi equations,
and in \S\ref{Tmnelectromagnetic}
the same condition, equation~(\ref{Apotzero}),
emerged from requiring that the
angular components of the electromagnetic energy-momentum be isotropic,
as conformal separability requires.

It has now been seen
that the potentials $\Apot_x \pm \Apot_t$
are sourced respectively by ingoing and outgoing collisionless streams,
\S\ref{collisionlesscurrentsource}.
Thus the condition
$\Apot_t = \pm \Apot_x$
requires that only one of the streams can be charged,
and the other must be neutral.
If the ingoing stream is charged, then $\Apot_t = \Apot_x$,
while if the outgoing stream is charged, then $\Apot_t = - \Apot_x$.

If the ingoing stream is charged, then $\Qelecbh^+$ is non-zero,
while if the outgoing stream is charged, then $\Qelecbh^-$ is non-zero.

In a real astronomical black hole,
collisions and magnetohydrodynamic processes are likely
to keep charged particles tightly coupled above the inner horizon,
forcing them into a common ingoing or outgoing stream
before inflation ignites.
Thus the condition that only one stream be charged
is physically realistic.

\subsection{Einstein and energy-momentum tensors}
\label{Einstein8}

For the solutions to be valid,
Einstein's equations must also be satisfied.
Equations~(124) of Paper~2 give 8 of the Einstein components
in the conformally stationary limit
(the remaining 2 components are considered in \S\ref{Einstein2}):
\begin{subequations}
\label{GUvcsimp}
\begin{align}
\label{GxtxUvsimp}
  \rho^2
  \left(
  {G_{xx} + G_{tt} \over 2}
  \,\pm\,
  G_{xt}
  \right)
  &=
  {\Ux \mp \vel
  \over - \Deltax}
  \left(
  \Deltax^\prime \pm \vel
  \right)
  \ ,
\\
\label{GxtyUvsimp}
  \rho^2
  \left(
  G_{xy}
  \,\pm\,
  G_{ty}
  \right)
  &=
  -
  {\Ux \mp \vel
  \over \sqrt{- \Deltax \Deltay}}
  \Deltay
  {\partial \ln \rhosep^2 \over \partial y}
  \ ,
\\
\label{GxtphiUvsimp}
  \rho^2
  \left(
  G_{x\phi}
  \,\pm\,
  G_{t\phi}
  \right)
  &=
  \pm
  {\Ux \mp \vel
  \over \sqrt{- \Deltax \Deltay}}
  \left(
  {\Deltay \over \sigma^2} {\dd \omegax \over \dd x}
  \mp
  2 \vel \omegay
  \right)
  \ ,
\\
\label{GyphiyUvsimp}
  \rho^2
  \left(
  {G_{yy} - G_{\phi\phi} \over 2}
  \,\pm\,
  \im
  G_{y\phi}
  \right)
  &=
  0
  \ ,
\end{align}
\end{subequations}
in which
%$\omegax^\prime \equiv \left. \dd \omegax / \dd x \right|_{\xin}$
$\Deltax^\prime \equiv \left. \dd \Deltax / \dd x \right|_{\xin}$
%are the derivatives of $\omegax$ and
is the (positive) derivative of
the electrovac horizon function $\Deltax$ at the inner horizon
$x = \xin$.

In the present situation,
there are two sources of energy-momentum,
electromagnetic and collisionless.
Given the expressions~(\ref{EiBelec}) for the radial electric and magnetic
fields $\Er$ and $\Br$,
and the solutions~(\ref{ApotQ}) for the potentials $\Apot_x \pm \Apot_t$,
and given that only one of the ingoing or outgoing streams can be charged
(which implies that $\Qelec^+ \Qelec^- = 0$),
equations~(\ref{Tmnem}) for the electromagnetic energy-momentum $\Te_{kl}$ are
\begin{subequations}
\label{TmnemU}
\begin{align}
\label{TxtemU}
  8\pi \rho^2
  \left(
  {\Te_{xx} + \Te_{tt} \over 2} \pm \Te_{xt}
  \right)
  &=
  {\Ux \mp \vel
  \over - \Deltax}
  X^\pm
  {\vel \omegay \over \Deltay}
  \ ,
\\
\label{TxyemU}
  8\pi \rho^2
  \left(
  \Te_{xy} \pm \Te_{ty}
  \right)
  &=
  \pm
  {\Ux \mp \vel
  \over \sqrt{-\Deltax \Deltay}}
  X^\pm
  {1 \over \sigma^2} {\dd \omegay \over \dd y}
  \ ,
\\
\label{TxphiemU}
  8\pi \rho^2
  \left(
  \Te_{x\phi} \pm \Te_{t\phi}
  \right)
  &=
  {\Ux \mp \vel
  \over \sqrt{-\Deltax \Deltay}}
  X^\pm
  %\left[
  {\partial \over \partial x}
  \ln
  \left(
  {1 \over \sigma^2} {\dd \omegax \over \dd x}
  \right)
  %+
  %z {\vel \omegay \over \Deltay}
  %\right]
  \ ,
\\
\label{TyphiemU}
  {\Te_{yy} - \Te_{\phi\phi} \over 2} \pm \im \Te_{y\phi}
  &=
  0
  \ ,
\end{align}
\end{subequations}
where
\begin{equation}
\label{Xpm}
  X^\pm
  \equiv
  {\vel \omegay \over \uel \mp \vel}
  \left[ {\Qelecbh^\pm \over 4 \rhosep ( f_0 g_1 + g_1 g_0 )}
  {\dd \omegax \over \dd x} \right]^2
  \ .
\end{equation}
The total energy-momentum
prescribed by the Einstein components~(\ref{GUvcsimp}),
minus the electromagnetic energy-momentum~(\ref{TmnemU}), is
\begin{subequations}
\label{TUvcsimp}
\begin{align}
\label{TxtxUvsimp}
  8\pi \rho^2
  \left(
  {T_{xx} + T_{tt} \over 2}
  \,\pm\,
  T_{xt}
  \right)
  &=
  {\Ux \mp \vel
  \over - \Deltax}
  \left(
  \Deltax^\prime \pm \vel
  -
  X^\pm
  {\vel \omegay \over \Deltay}
  \right)
  \ ,
\\
\label{TxtyUvsimp}
  8\pi \rho^2
  \left(
  T_{xy}
  \,\pm\,
  T_{ty}
  \right)
  &=
  -
  {\Ux \mp \vel
  \over \sqrt{- \Deltax \Deltay}}
  \left(
  \Deltay
  {\partial \ln \rhosep^2 \over \partial y}
  \pm
  X^\pm
  {1 \over \sigma^2} {\dd \omegay \over \dd y}
  \right)
  \ ,
\\
\label{TxtphiUvsimp}
  8\pi \rho^2
  \left(
  T_{x\phi}
  \,\pm\,
  T_{t\phi}
  \right)
  &=
  \pm
  {\Ux \mp \vel
  \over \sqrt{- \Deltax \Deltay}}
  \left[
  \Deltay
  {1 \over \sigma^2}
  {\dd \omegax \over \dd x}
  \mp
  2 \vel \omegay
  \mp
  X^\pm
  {\partial \over \partial x}
  \ln
  \left(
  {1 \over \sigma^2} {\dd \omegax \over \dd x}
  \right)
  \right]
  \ ,
\\
\label{TyphiyUvsimp}
  8\pi \rho^2
  \left(
  {T_{yy} - T_{\phi\phi} \over 2}
  \,\pm\,
  \im
  T_{y\phi}
  \right)
  &=
  0
  \ .
\end{align}
\end{subequations}
For the solution to be consistent,
the energy-momentum tensor given by equations~(\ref{TUvcsimp})
must be consistent with being sourced by collisionless streams.
Indeed it is, as shown in the next subsection.

\subsection{Collisionless source of energy-momentum}
\label{collisionlessenergymomentumsource}

The energy-momentum tensor~(\ref{TUvcsimp}) coincides with that of
a sum of ingoing ($+$) and outgoing ($-$) collisionless streams
\begin{equation}
\label{Tpm}
  T_{kl}
  =
  N^+ p^+_k p^+_l
  +
  N^- p^-_k p^-_l
  \ ,
\end{equation}
with number densities
\begin{equation}
\label{Npm}
  N^\pm
  =
  {1 \over 16\pi}
  ( \Ux \mp \vel )
  \left(
  \Deltax^\prime \pm \vel
  -
  X^\pm
  {\vel \omegay \over \Deltay}
  \right)
  \ ,
\end{equation}
and tetrad-frame momenta
\begin{equation}
\label{ppm}
  p^\pm_k
  =
  {1 \over \rho}
  \left\{
  -
  {1 \over \sqrt{- \Deltax}}
  \, , \ 
  \mp
  {1 \over \sqrt{- \Deltax}}
  \, , \ 
  {1 \over \sqrt{\Deltay}}
  {
  \displaystyle
  \Deltay
  {\partial \ln \rhosep^2 \over \partial y}
  \pm
  X^\pm
  {1 \over \sigma^2} {\dd \omegay \over \dd y}
  \over
  \displaystyle
  \Deltax^\prime \pm \vel
  -
  X^\pm
  {\vel \omegay \over \Deltay}
  }
  \, , \ 
  \mp
  {1 \over \sqrt{\Deltay}}
  {
  \displaystyle
  \Deltay
  {1 \over \sigma^2} {\dd \omegax \over \dd x}
  \mp
  2 \vel \omegay
  \mp
  X^\pm
  {\partial \over \partial x}
  \ln
  \left(
  {1 \over \sigma^2} {\dd \omegax \over \dd x}
  \right)
  \over
  \displaystyle
  \Deltax^\prime \pm \vel
  -
  X^\pm
  {\vel \omegay \over \Deltay}
  }
  \right\}
  \ ,
\end{equation}
as long as condition~(\ref{Dsmall}) on the horizon function $\Deltax$ holds.
The densities~(\ref{Npm}) and momenta~(\ref{ppm})
are defined up to arbitrary normalization factors
such that the energy-momentum~(\ref{Tpm}) is fixed.
The ingoing and outgoing densities $N^\pm$, equation~(\ref{Npm}),
conform to the Hamilton-Jacobi behaviour~(\ref{N}),
satisfying (see equation~(117) of Paper~2)
\begin{equation}
  N^\pm
  \propto
  {1 \over \rho^2}
  \propto
  \Ux \mp \vel
  \ ,
\end{equation}
the remaining factors in equation~(\ref{Npm})
being constant along the path of a stream.
%While the charge density $q^\pm N^\pm$ and the density $N^\pm$
%both vary as $1/\rho^2$ along the path of the stream,
%they have different functional dependences on conformal time $t$ and radius $x$:
%the charge density depends on conformal time as $q^\pm N^\pm \ee^{\vel t}$,
%whereas the density $N^\pm$ is independnent of conformal time $t$.
Similarly the tetrad-frame momentum $p^\pm_k$, equation~(\ref{ppm}),
accords with the Hamilton-Jacobi form~(\ref{pktetrad}),
with constant Hamilton-Jacobi parameters $P_k$
along the path of the stream.

If the black hole were uncharged, then $X^\pm$ defined by equation~(\ref{Xpm})
would vanish,
and equations~(\ref{Npm}) and (\ref{ppm})
would reduce to equations~(127) and (128) of Paper~2.
If the black hole's charge-to-mass ratio $\Qelecbh / \Mbh$
is of order unity, then (one of, if only one stream is charged)
$X^\pm$ is of order unity,
but if the black hole's charge is small,
then $X^\pm$ too will be small.
In any case,
the charge of the black hole has little effect on
the collisionless densities $N^\pm$, equation~(\ref{Npm}),
the term proportional to $X^\pm$ being of order $\vel$
compared to the principal term $\Deltax^\prime$.
This expresses the fact that the radial components of the energy-momentum
are dominated by the streaming energy-momentum,
not the electromagnetic energy-momentum.
This in turn reflects the fact that the inflationary instability
is fundamentally gravitational, not electromagnetic.
However,
if the black hole's charge-to-mass ratio is of order unity,
then it has order unity effect
on the angular components of the collisionless momenta $p^\pm_k$,
equation~(\ref{ppm}).

The mean charge $\langle q^\pm \rangle$
per accreted particle is the ratio
of the charge density $q^\pm N^\pm$, equation~(\ref{qNpm}),
to the number density $N^\pm$, equation~(\ref{Npm}):
\begin{equation}
\label{qpm}
  \langle q^\pm \rangle
  \equiv
  {q^\pm N^\pm \over N^\pm}
  =
  \ee^{- \vel t}
  {\vel \Qelecbh^\pm \over 2 \rhosep^2 ( f_0 g_1 + f_1 g_0 )}
  {
  \displaystyle
  {\dd \omegax \over \dd x}
  \left[
  {\partial \over \partial x}
  \ln \left( {1 \over \sigma^2} {\dd \omegax \over \dd x} \right)
  \mp
  {\vel \omegay^2 \over \Deltay}
  \right]
  \over
  \displaystyle
  \Deltax^\prime \pm \vel
  -
  X^\pm
  {\vel \omegay \over \Deltay}
  }
  \ .
\end{equation}
The mean charge per particle~(\ref{qpm})
decreases with time as the black hole expands,
consistent with dimensional analysis:
\begin{equation}
\label{qevt}
  \langle q^\pm \rangle
  \propto
  \ee^{- \vel t}
  \ .
\end{equation}
This contrasts with the number densities $N^\pm$,
which are dimensionless, independent of conformal time $t$.
Aside from the dependence on $t$,
the remaining factors in the mean charge per particle~(\ref{qpm})
are just functions of latitude $y$,
the dependence on radius $x$ being frozen at its inner horizon value.

The angular components of the number-weighted momenta $p^\pm_k$
given by equation~(\ref{ppm})
differ from those of the charge-weighted momenta $p^\pm_k$
given by equation~(\ref{qppm}).
The difference poses no great difficulty,
but it does mean that the angular conditions on
the current and the energy-momentum of the charged stream
cannot be accomplished
simultaneously with a single collisionless component.
Two components to the charged stream would suffice.
For example, one component could be charged, fulfilling the
conditions~(\ref{qNpm}) and (\ref{qppm}),
and the other could be neutral,
its number and momentum chosen such that,
when added to those of the charged component,
their sum fulfills the conditions~(\ref{Npm}) and (\ref{ppm}).
More generally, both components could be charged,
the more highly charged stream providing much of the current,
and the more lightly charge stream providing much of the number density.

\subsection{Remaining Einstein components}
\label{Einstein2}

The remaining 2 Einstein components,
besides the 8 given as equations~(\ref{GUvcsimp}),
are $G_{xx} - G_{tt}$ and $G_{yy} + G_{\phi\phi}$.
These two components govern the evolution of the horizon function $\Deltax$
and inflationary exponent $\expinf$,
as described in \S{VIII\,D,E} of Paper~2.
Unlike other Einstein components,
these 2 Einstein components do not grow during inflation,
although they do grow during collapse when the conformal factor shrinks.
They do not grow firstly because
the trace of the collisionless energy-momentum remains negligible,
and the trace of the electromagnetic energy-momentum is zero,
and secondly because
the combination $G_{yy} + G_{\phi\phi}$ of angular components
of the energy-momentum is,
for both collisionless and electromagnetic contributions,
independent of the radial horizon function $\Deltax$, and therefore
they do not grow large at the inner horizon where $\Deltax \rightarrow -0$,
unlike radial components which are proportional to inverse powers of $\Deltax$.

To dominant order,
the evolution of the horizon function and inflationary exponent
during inflation and collapse
are unaffected by the collisionless and electromagnetic energy-momentum,
except that the initial electrovac solution sets
the derivative $\Deltax^\prime$ of the horizon function at the inner horizon.

The effect of the sub-dominant purely angular components of the
energy-momentum may be taken into account by solving the Einstein
equations for the 2 components to next higher order in $\Deltax / \Ux$,
as described in \S{VIII\,J} of Paper~2.
Given the expressions~(\ref{EiBelec}) for the radial electric and magnetic
fields $\Er$ and $\Br$,
and the solutions~(\ref{ApotQ}) for the potentials $\Apot_x \pm \Apot_t$,
the relevant components~(\ref{Txxmttem}) of
the electromagnetic energy-momentum $\Te_{kl}$ are
\begin{align}
\label{TxxmttemU}
  8\pi \rho^2
  \left(
  \Te_{xx} - \Te_{tt}
  \right)
  &=
  8\pi \rho^2
  \left(
  \Te_{yy} + \Te_{\phi\phi}
  \right)
  \frac{1}{2}
  \left[
  ( \Ux - \vel ) {X^+ \over \vel \omegay}
  +
  ( \Ux + \vel ) {X^- \over \vel \omegay}
  \right]
  \left[
  \left(
  {\partial \over \partial x}
  \ln
  \left(
  {1 \over \sigma^2} {\dd \omegax \over \dd x}
  \right)
  %+
  %z {\vel \omegay \over \Deltay}
  \right)^2
  +
  \left(
  {1 \over \sigma^2} {\dd \omegay \over \dd y}
  \right)^2
  \right]
  \ .
\end{align}
If the charge-to-mass ratio of the black hole is of order unity,
so that one of $X^\pm$ is of order unity,
then the electromagnetic energy-momentum~(\ref{TxxmttemU}),
which is of order $\sim \Ux / \vel$,
is larger by of order $1 / \vel$
than the corresponding collisionless energy-momentum
given by equation~(136) of Paper~2.
Thus the angular components of the electromagnetic energy-momentum,
though still a sub-dominant influence
on the evolution of the horizon function and inflationary exponent
during inflation and collapse,
have a potentially larger influence than the
angular components of the collisionless energy-momentum.

The electromagnetic energy-momentum components~(\ref{TxxmttemU})
can be taken into account in the Einstein equations by
introducing source functions $F_X$ and $F_Y$
to the evolutionary equations for $\Ux$ and $\Deltax$,
equations~(139) of Paper~2.
The relevant source functions are
\begin{equation}
\label{FXYem}
  F_X
  =
  F_Y
  =
  {4\pi \rho^2
  \left(
  \Te_{yy} + \Te_{\phi\phi}
  \right)
  \over \Ux}
  \ .
\end{equation}
During early inflation,
while $\Ux$ remains at its initial value of $\uel$,
the source functions are just those of the electrovac spacetime,
which are already taken into account in the solution.
Once $\Ux$ has increased appreciably above $\uel$,
the horizon function is already exponentially tiny,
$| \Deltax | \sim \ee^{- 1 / \vel}$.
The condition that the sources $F_X$ and $F_Y$ have negligible
influence on the evolution of $\Ux$ and $\Deltax$
is that $\Ux F_X \ll \Ux^2 / | \Deltax |$,
which for the source functions~(\ref{FXYem}) is equivalent to
$| \Deltax | / \Ux \ll 1 / \vel$.
This is well-satisfied during inflation because
$| \Deltax |$ is exponentially tiny,
and during collapse because $| \Deltax |$ remains small
and $\Ux$ grows exponentially large.

Eventually the conformally separable solution breaks down,
but not because of the electromagnetic energy-momentum
(provided that only one of the ingoing and outgoing streams is charged).
Rather, when the angular motion of the collisionless
streams becomes comparable to their radial motion,
which happens when $| \Deltax | \gtrsim 1$ at the end of collapse,
the conformally separable Einstein and Maxwell equations
cease to be satisfied by the energy-momentum and current of
freely-falling collisionless streams.

\section{Boundary conditions}

Maxwell's and Einstein's equations have prescribed the
forms of the electric current~(\ref{jpm})--(\ref{qppm})
and energy-momentum~(\ref{Tpm})--(\ref{ppm})
of the freely-falling ingoing and outgoing collisionless streams
during inflation and collapse.
These forms are required by the condition of conformal separability.
As in Paper~2,
because the accretion rate is asymptotically tiny,
the charge and energy-momentum of the collisionless streams
are negligible above the inner horizon,
so have negligible effect on the geometry above the inner horizon.
From the perspective of boundary conditions,
what is important is the form of the collisionless streams
incident on the inner horizon.

During inflation and collapse,
%ingoing and outgoing collisionless streams stream hyper-relativistically
%through each other along the principal ingoing and outgoing null directions.
the radial ($x$-$t$) components of the momenta
$p^\pm_k$ of the streams
are of order
$1/\sqrt{- \Deltax}$ times the angular ($y$-$\phi$) components,
and dominate as long as $| \Deltax | \ll 1$.
The solution for the dominant radial components is essentially unaffected
by the angular components.
If only the dominant radial Maxwell and Einstein equations
are required to be satisfied,
then only radial boundary conditions, \S\ref{radialbcs},
need be imposed.
If the sub-dominant angular Maxwell and Einstein equations
are required to be satisfied,
then also angular boundary conditions, \S\ref{angularbcs},
must be imposed.
If the sub-sub-dominant purely angular Einstein equations
are further required to be satisfied,
then yet further boundary conditions, \S\ref{angularangularbcs},
must be imposed.

\subsection{Density and charge of collisionless streams incident on the inner horizon}
\label{radialbcs}

The indispensable boundary conditions are those on the radial ($x$-$t$)
components of the collisionless current and energy-momentum
incident on the inner horizon.

The ingoing ($+$) and outgoing ($-$) radial energy-momenta
during inflation and collapse are,
equations~(\ref{Tpm})--(\ref{ppm}),
\begin{equation}
  T^\pm_{xx}
  =
  \pm
  T^\pm_{xt}
  =
  T^\pm_{xx}
  =
  {N^\pm \over \rho^2 | \Deltax |}
  \ .
\end{equation}
The initial values of these components are set by the incident
number densities $N^\pm$, equation~(\ref{Npm}), which are,
since $\Ux = \uel$ initially,
\begin{equation}
\label{Npminit}
  N^\pm
  =
  {1 \over 16\pi}
  ( \uel \mp \vel )
  \left(
  \Deltax^\prime \pm \vel
  -
  X^\pm
  {\vel \omegay \over \Deltay}
  \right)
  \ .
\end{equation}
If the sub-dominant parts proportional to $\vel$ are neglected,
then the density~(\ref{Npminit}) is uniform, independent of latitude,
meaning that the required accretion flow is monopole.
If the sub-dominant contributions proportional to $\vel$ are
taken into account, which is necessary if the boundary conditions
are to be imposed to angular order, \S\ref{angularbcs},
then the part proportional to $X^\pm$,
which arises from the presence of the electromagnetic field,
introduces a small angular dependence of the incident densities $N^\pm$.

The ingoing ($+$) and outgoing ($-$) radial currents
during inflation and collapse are,
equations~(\ref{jpm})--(\ref{qppm}),
\begin{equation}
\label{jpmxt}
  j^\pm_x = \pm j^\pm_t
  =
  {q^\pm N^\pm \over \rho \sqrt{-\Deltax}}
  \ .
\end{equation}
%Recall that only one of the ingoing and outgoing streams can be charged,
%so one of these currents is zero.
The initial values of the radial currents~(\ref{jpmxt})
are set by the incident charge densities $q^\pm N^\pm$,
equation~(\ref{qNpm}),
which are,
\begin{equation}
\label{qNpminit}
  q^\pm N^\pm
  =
  {\vel \Qelecbh^\pm \over 32\pi \rhosep^2 ( f_0 g_1 + f_1 g_0 )}
  {\dd \omegax \over \dd x}
  \left[
  {\partial \over \partial x}
  \ln \left( {1 \over \sigma^2} {\dd \omegax \over \dd x} \right)
  \mp
  {\vel \omegay^2 \over \Deltay}
  \right]
  \ .
\end{equation}
Unlike the number density $N^\pm$,
the charge density $q^\pm N^\pm$ varies significantly with latitude.
If the sub-dominant term proportional to $\vel$ inside square brackets
on the right hand side of equation~(\ref{qNpminit}) is neglected,
then the incident charge density
is proportional to the radial electric field $E$,
\begin{equation}
\label{qNpminitE}
  q^\pm N^\pm
  =
  {\vel E \over 4\pi}
  \ .
\end{equation}
The different angular behaviours of
the number and charge and densities~(\ref{Npminit}) and (\ref{qNpminit})
mean that the mean charge per particle must vary with latitude,
equation~(\ref{qpm}).

The charge density equation~(\ref{qNpminit})
relates the ingoing and outgoing cumulative black hole charges $\Qelecbh^\pm$
to the rates $q^\pm N^\pm$ at which ingoing and outgoing charges are accreted.
This is a feature of conformally time-translation symmetric (self-similar)
spacetimes,
that cumulative properties are determined by the rate of their accretion.

\subsection{Angular motions of collisionless streams incident on the inner horizon}
\label{angularbcs}

The angular components of the momenta of the collisionless streams
are sub-dominant during inflation and collapse.
If the sub-dominant Einstein and Maxwell equations are to be satisfied,
then angular boundary conditions must be imposed
in addition to the radial boundary conditions.

Equation~(\ref{ppm}) specifies the required
number-weighted
tetrad-frame momentum $p^\pm_k$
of the ingoing and outgoing streams.
The corresponding number-weighted Hamilton-Jacobi parameters $P^\pm_k$ are
\begin{equation}
\label{Ppm}
  P^\pm_k
  =
  \left\{
  -
  1
  \, , \ 
  \mp
  1
  \, , \ 
  {
  \displaystyle
  \Deltay
  {\partial \ln \rhosep^2 \over \partial y}
  \pm
  X^\pm
  {1 \over \sigma^2} {\dd \omegay \over \dd y}
  \over
  \displaystyle
  \Deltax^\prime \pm \vel
  -
  X^\pm
  {\vel \omegay \over \Deltay}
  }
  \, , \ 
  \mp
  {
  \displaystyle
  \Deltay
  {1 \over \sigma^2} {\dd \omegax \over \dd x}
  \mp
  2 \vel \omegay
  \mp
  X^\pm
  {\partial \over \partial x}
  \ln
  \left(
  {1 \over \sigma^2} {\dd \omegax \over \dd x}
  \right)
  \over
  \displaystyle
  \Deltax^\prime \pm \vel
  -
  X^\pm
  {\vel \omegay \over \Deltay}
  }
  \right\}
  \ .
\end{equation}
Similarly, equation~(\ref{qppm}) specifies the required
charge-weighted
tetrad-frame momentum $p^\pm_k$ of whichever of the ingoing
or outgoing streams is charged,
and the corresponding
charge-weighted Hamilton-Jacobi parameters $P^\pm_k$ are
\begin{equation}
\label{qPpm}
  P^\pm_k
  =
  \left\{
  -
  1
  \, , \ 
  \mp
  1
  \, , \ 
  {
  \displaystyle
  {1 \over \sigma^2} {\dd \omegay \over \dd y}
  \over
  \displaystyle
  {\partial \over \partial x}
  \ln \left( {1 \over \sigma^2} {\dd \omegax \over \dd x} \right)
  \mp
  {\vel \omegay^2 \over \Deltay}}
  \, , \ 
  \pm
  {
  \displaystyle
  {\partial \over \partial x}
  \ln
  \left(
  {1 \over \sigma^2} {\dd \omegax \over \dd x}
  \right)
  \over
  \displaystyle
  {\partial \over \partial x}
  \ln \left( {1 \over \sigma^2} {\dd \omegax \over \dd x} \right)
  \mp
  {\vel \omegay^2 \over \Deltay}
  }
  \right\}
\ .
\end{equation}
The angular components $\Py^\pm$ and $P^\pm_\phi$ vary with latitude $y$,
but remain frozen at their inner horizon values along the path
of a stream during inflation and collapse, as long as $| \Deltax | \ll 1$,
that is, until angular motions become important at the end of collapse,
when $| \Deltax | \gtrsim 1$.

The number- and charge-weighted angular motions~(\ref{Ppm}) and (\ref{qPpm})
differ.
To achieve both angular conditions requires that the charged stream
contain more than one component.
As remarked in \S\ref{collisionlessenergymomentumsource},
both conditions~(\ref{Ppm}) and (\ref{qPpm}) together can be accomplished
with two (or more) components,
a more highly charged component that produces most of the current,
and a more lightly charged component that produces most of the number density.

\subsection{Feasibility of the angular boundary conditions with collisionless streams accreted from outside the outer horizon}

Can the boundary conditions~(\ref{Ppm}) and (\ref{qPpm})
on the angular motions of incident ingoing and outgoing streams
be accomplished by real collisionless streams?
In \S{X\,C} of Paper~2 \cite{Hamilton:2010b}
it was shown that, in the case of an uncharged black hole,
the conditions cannot be accomplished if the collisionless streams are
required to be accreted from outside the outer horizon.
In the present subsection it is shown that electric charge
alleviates the problem, but does not eliminate it.

If every particle is accreted from outside the outer horizon,
then the Hamilton-Jacobi parameter
$P_t$
must necessarily be negative (ingoing) at the outer horizon
for every particle.
For a particle of charge $q$,
the $P_t$ and $P_\phi$ of the particle at the inner horizon must then satisfy
the inequality, generalizing inequality~(161) of Paper~2,
\begin{equation}
\label{Ptconditionq}
  P_t
  \,\leq\,
  P_\phi
  {\omegaxin - \omegaxout \over 1 - \omegaxout \omegayin}
  +
  q
  \left(
  -
  \Apot_{t, {\rm in}}
  %+
  %{\omegaxin - \omegaxout \over 1 - \omegaxout \omegayin}
  %\Apot_{\phi , {\rm in}}
  +
  {1 - \omegaxin \omegayin \over 1 - \omegaxout \omegayin}
  \Apot_{t, {\rm out}}
  \right)
  \ ,
\end{equation}
where subscripts ${\rm out}$ and ${\rm in}$
denote values at the outer and inner horizons.
If the particle has the same charge as the black hole,
as is more likely
since the black hole inherits its charge from the accreted streams,
then the charge-dependent factor in condition~(\ref{Ptconditionq})
is positive,
so that the allowed region of the $P_t$--$P_\phi$ plane
includes a finite region around the origin $P_t = P_\phi = 0$.
Physically,
the black hole's charge repels the charged particle
between the outer and inner horizon,
increasing its $P_t$.
This is sufficient to allow the ratio $P_t / P_\phi$
prescribed by either of equations~(\ref{Ppm}) and (\ref{qPpm})
to be accomplished at all latitudes.

If both ingoing and outgoing streams were permitted to be charged,
with the same charge as the black hole,
then condition~(\ref{Ptconditionq})
could be accomplished by both streams at all latitudes.
If both streams are charged, however,
the spacetime does not remain conformally separable
because the diagonal angular components
of the electromagnetic energy-momentum diverge near the inner horizon.
Conformal separability begins to break down when the diverging component
$\rho^2 ( T^{\rm e}_{yy} - T^{\rm e}_{\phi\phi} )$,
equation~(\ref{Typhiem}),
which is dimensionless,
%(independent of conformal time $t$)
is of order unity, which happens when
\begin{equation}
\label{Dxnosep}
  | \Deltax |
  \sim
  {\vel^2 \Qelecbh^+ \Qelecbh^-
  \over \rho^2}
  \ .
\end{equation}
This is small,
both because it is proportional to the square $\vel^2$
of the small accretion rate $\vel$,
and because real black holes are likely to have small charge,
so $\Qelecbh^\pm$ will be small.
The black hole charge can, despite its smallness,
affect charged particles because of the large charge-to-mass ratio
of real particles, protons and electrons.
Although the value~(\ref{Dxnosep})
of the horizon function $\Deltax$ when conformal separability fails is small,
it is nonetheless large compared
to the exponentially tiny values to which the horizon function
is driven during inflation.

If conformal separability is required to persist during inflation and
collapse, then one or other of the
ingoing and outgoing streams must be neutral.
For the neutral stream,
as discussed in \S{X\,C} of Paper~2,
the condition~(\ref{Ptconditionq})
(with $q = 0$)
cannot be satisfied simultaneously at all latitudes.

%In a real black hole,
%if only one stream is charged,
%then it is the outgoing stream that is more likely to be charged,
%necessarily with the same charge as the black hole
%since the black hole inherits its charge from the stream.
%Particles with the same sign charge as the black hole
%are repelled by the charge of the black hole,
%and tend to become outgoing.

To summarize,
if conformal separability to angular order is demanded
only during the early part of inflation,
then the required conditions on the incident angular motions
of ingoing and outgoing streams
can be accomplished by collisionless streams accreted entirely
from outside the outer horizon,
provided that both streams are charged,
with the same charge as the black hole.
However, if conformal separability is demanded
throughout inflation and collapse,
then one of the streams must be neutral,
and the angular conditions on that neutral stream
cannot be accomplished by a collisionless stream accreted
entirely from outside the outer horizon.

\subsection{Dispersion of angular motions incident on the inner horizon}
\label{angularangularbcs}

The purely angular components of the energy-momentum tensor
are sub-sub-dominant during inflation and collapse while $| \Deltax | \ll 1$.
Their effect can nevertheless be taken into account by solving
the Einstein equations to next higher order in $\Deltax / \Ux$,
as described in \S\ref{Einstein2}.
If the Einstein equations are required to hold to this order,
then the condition of conformal separability requires that
the $2 \times 2$ angular submatrix of the energy-momentum tensor
must be isotropic.
The electromagnetic energy-momentum tensor~(\ref{Tmnem})
satisfies the condition of isotropy
provided that condition~(\ref{Apotzero}) holds,
which is true provided that
only one of the ingoing or outgoing streams is charged.
As discussed in \S{X\,D} of Paper~2,
the condition of angular isotropy on the collisionless energy-momentum
can also be contrived,
by allowing multiple incident ingoing and outgoing streams
whose angular components of momentum satisfy~(\ref{Ppm}) in the mean,
but are isotropic in their mean squares.

The more precisely
the Einstein and Maxwell equations are required to be satisfied,
the more special and contrived the conditions
required by conformal separability become.

\section{Conclusions}
\label{conclusions}

The conformally stationary, axisymmetric, conformally separable solutions
for the interior structure of rotating black holes
found in Paper~2 \cite{Hamilton:2010b}
generalize to the case of charged rotating black holes.
Maxwell's equations separate consistently with Einstein's equations.
The collisionless fluid accreted
by the black hole is permitted to be electrically charged,
and the charge of the black hole is produced self-consistently
by the accretion of charge.
As in the uncharged case,
hyper-relativistic counter-streaming between ingoing and outgoing streams
drives mass inflation at the inner horizon, followed by collapse.

The only anomaly is that conformal separability in charged black holes
requires that only one of the ingoing or outgoing streams
can be charged: the other stream must be neutral.
If both streams are charged, then
conformal separability holds during early inflation,
but radially counter-streaming electric currents,
in concert with the rotation of the black hole,
generate angular electromagnetic fields
that cause the non-isotropic diagonal angular
component of the electromagnetic energy-momentum to diverge,
destroying conformal separability.
I suspect that the physical reason for the breakdown of conformal separability
is that if both streams are charged, then they can exchange
energy-momentum via the electromagnetic field that they mutually create.
This breaks the $\vel \leftrightarrow - \vel$ symmetry
between the ingoing and outgoing streams,
which appears to be important to the existence of the solutions.

In practice,
the condition that only one stream be charged is physically realistic,
since collisions and magnetohydrodynamic processes are likely
to keep charged particles tightly coupled above the inner horizon,
forcing them into a common ingoing or outgoing stream
before inflation ignites.

The most important equations in this paper are the Maxwell
equations~(\ref{MaxwellZ}).
These equations hold over the entire regime of interest,
from electrovac through inflation and collapse.
In concert with the Einstein equations~(88) from Paper~2,
their solution yields the full suite of both
stationary, separable electrovac,
and conformally stationary, conformally separable inflationary solutions.

The condition of conformal separability imposes a hierarchy
of boundary conditions on the collisionless streams
incident on the inner horizon.
The indispensible boundary condition is on the dominant radial components
of the collisionless energy-momentum and current.
The radial conditions require that
the incident number densities $N^\pm$, equation~(\ref{Npminit}),
of the ingoing and outgoing streams
must be uniform with latitude
(with a sub-dominant order $\vel$ angular dependence
arising from the electromagnetic field).
By contrast the incident charge densities $q^\pm N^\pm$,
equation~(\ref{qNpminit})
must vary with latitude with, to leading order,
the same angular dependence as the radial electric field $E$,
equation~(\ref{qNpminitE}).
The different angular dependences of the number and charge densities
$N^\pm$ and $q^\pm N^\pm$
imply that the incident mean charge per particle must vary with latitude,
equation~(\ref{qpm}).

If the sub-dominant radial-angular components of the Einstein equations
and angular components of the Maxwell equations
are required to be satisfied,
then conformal separability requires that the angular components
of the number-weighted and charge-weighted momenta of the incident streams
have Hamilton-Jacobi parameters $P_k$ satisfying equations~(\ref{Ppm})
and (\ref{qPpm}) respectively.
The number- and charge-weighted angular motions differ,
implying that the conditions cannot be accomplished by a
collisionless charged stream containing just one component.
However, the angular conditions on a charged stream
can be achieved with two (or more) components,
a more highly charged component that produces most of the current,
and a more lightly charged component that produces most of the number density.

In Paper~2, it was emphasized that the solutions had the limitation
that the angular conditions on the incident ingoing and outgoing streams
could not be achieved
by collisionless streams that fall freely from outside the outer horizon.
The present paper finds that the required angular conditions
can be achieved by a charged stream,
provided that the stream has the same sign charge as the black hole,
but not by a neutral stream.
As commented above,
if both streams are charged, then conformal separability
holds only during early inflation.
If conformal separability is required to hold throughout inflation and collapse,
then one of the streams must be neutral,
and then the angular conditions on the incident neutral stream
%imposed by conformal separability to angular order
cannot be achieved by a stream that falls freely from outside the outer horizon.

If the sub-sub-dominant purely angular Einstein equations are required to
be satisfied,
then conformal separability requires that the angular energy-momentum
tensor be isotropic
(proportional to the unit $2 \times 2$ matrix).
The collisionless energy-momentum can be contrived to be isotropic,
and the electromagnetic energy-momentum is isotropic provided that only
one of the ingoing or outgoing streams is charged.

\begin{acknowledgements}
I thank Gavin Polhemus for numerous conversations
that contributed materially to the development of the ideas herein.
This work was supported by NSF award
AST-0708607.
\end{acknowledgements}

\section*{References}

\bibliographystyle{unsrt}
\bibliography{bh}

\appendix

\section{Non-vanishing $Z_x$ and $Z_t$}
\label{Zsec}

The calculation of the evolution of the potentials $\Apot_x \pm \Apot_t$
in \S\ref{evolutionA}
was premised in part on the vanishing of $Z_x$ and $Z_t$
defined by equations~(\ref{Zk}).
This Appendix examines what happens if $Z_x$ and $Z_t$ do not vanish.

The conclusion is that $Z_x$ and $Z_t$ must take a certain form~(\ref{Zpmz})
in order that Maxwell's equations can continue to be satisfied
by the current of collisionless streams.
By adjusting the proportionality factor $z$ in this form~(\ref{Zpmz}),
the angular current $j^\pm_\phi$
of the collisionless ingoing or outgoing stream
can be adjusted to be an essentially arbitrary function of latitude $y$.
This change relaxes the angular conditions on the current
imposed by the assumption of conformal separability,
but leaves all the conclusions of the main text unchanged.
Only the azimuthal angular current $j^\pm_\phi$ is adjustable:
the radial currents $j^\pm_x$ and $j^\pm_t$ are scarcely affected,
and the angular current $j^\pm_y$ is not affected at all.

\subsection{Required form of $Z_x$ and $Z_t$}
\label{Zform}

Currents sourced by Maxwell's equations~(\ref{MaxwellZnomag})
with non-vanishing $Z_x$ and $Z_t$ must continue
to fit the form of collisionless currents.
For the radial currents $j_x \pm j_t$,
this requires that the set of terms proportional to $Z_t$
on the right hand side of equation~(\ref{MaxwellZnomagxt})
must be proportional to $\Apot_x \pm \Apot_t$
times an appropriate factor of the horizon function $\Deltax$
(the sub-dominant third term inside square brackets
is temporarily neglected in this subsection \S\ref{Zform},
but is reinstated thereafter):
\begin{equation}
\label{Zeq}
  \left(
  {\partial \over \partial x}
  \mp
  {\vel \over \Deltax}
  \right)
  Z_t
  \propto
  {\Apot_x \pm \Apot_t \over \Deltax}
  \ .
\end{equation}
Equation~(\ref{Zeq}) requires that
\begin{equation}
  {\partial Z_t \over \partial x}
  =
  -
  {\Apot_x \over \Apot_t}
  {\vel \over \Deltax}
  Z_t
  \ ,
\end{equation}
which,
given that $\Apot_x = \pm \Apot_t$,
integrates to
%\begin{equation}
%  \ln Z_t
%  =
%  \mbox{constant}
%  -
%  \int {\Apot_x \over \Apot_t} {\vel \over \Deltax} \dd x
%  =
%  \mbox{constant}
%  -
%  \frac{1}{4}
%  \int {\Apot_x \over \Apot_t}
%  \dd \ln \left( {\Ux + \vel \over \Ux - \vel} \right)
%\end{equation}
\begin{equation}
\label{ZtA}
  Z_t
  \propto
  \Apot_x \pm \Apot_t
  \ ,
\end{equation}
whichever one of $\Apot_x \pm \Apot_t$ is non-vanishing.
Equation~(\ref{dApot}) shows that $Z_x \pm Z_t$
is a source for the evolution of
$\Apot_x \pm \Apot_t$.
To ensure that the combination
$\Apot_x \pm \Apot_t$
that vanishes continues to vanish as $| \Deltax |$ decreases
to exponentially tiny values,
the corresponding source $Z_x \pm Z_t$ must also vanish.
Putting this condition together with~(\ref{ZtA}) requires that
\begin{equation}
\label{ZxtA}
  Z_x
  =
  \pm
  Z_t
  \propto
  \Apot_x \pm \Apot_t
  \ ,
\end{equation}
for whichever one of $\Apot_x \pm \Apot_t$ is non-vanishing.
The constraints~(\ref{ZxtA}) on $Z_x$ and $Z_t$ are conveniently written
\begin{equation}
\label{Zpmz}
  Z_x \pm Z_t
  =
  z
  {\vel \omegay \over \Deltay}
  \left( \Apot_x \pm \Apot_t \right)
  \ ,
\end{equation}
for some factor $z$, which could be an arbitrary function of angle $y$.

\subsection{Evolution of the electromagnetic potential and enclosed charge}

Inserting the result~(\ref{Zpmz}) for $Z_x \pm Z_t$
into the evolutionary equation~(\ref{dApot}) for $\Apot_x \pm \Apot_t$
gives
\begin{equation}
\label{dApotz}
  \left(
  {\partial \over \partial x}
  \pm
  {\vel \over \Deltax}
  -
  z
  {\vel \omegay \over \Deltay}
  \right)
  {\Apot_x \pm \Apot_t \over \dd \omegax / \dd x}
  =
  0
  \ .
\end{equation}
The driving term proportional to $\vel / \Deltax$,
which diverges at the inner horizon $\Deltax \rightarrow -0$,
dominates the term proportional to $z$
in the conformally stationary limit.
The $z$-term changes the evolution of the potentials negligibly.

\subsection{Maxwell's equations}

Given $Z_x$ and $Z_t$ from equation~(\ref{Zpmz}),
Maxwell's equations~(\ref{MaxwellZnomag}) become equations~(\ref{jk})
with, in the same format as equations~(\ref{MaxwellZzero}),
\begin{subequations}
\label{MaxwellZz}
\begin{align}
\label{MaxwellZzxt}
  j^\pm_x
  =
  \pm
  j^\pm_t
  &\equiv
  {1 \over 8\pi \rho^3}
  {\vel \over \sqrt{- \Deltax}}
  ( \Apot_x \pm \Apot_t )
  \left\{
  \pm \,
  {\partial \over \partial x}
  \ln \left( {1 \over \sigma^2} {\dd \omegax \over \dd x} \right)
  -
  {\vel \omegay^2 \over \Deltay}
  +
  z
  {\omegay \over \Deltay}
  \left[
  \pm \vel
  -
  \Deltax
  \left(
  {\partial \over \partial x}
  \ln
  \left(
  {1 \over \sigma}
  {\dd \omegax \over \dd x}
  \right)
  +
  {z \over 2}
  {\vel \omegay \over \Deltay}
  \right)
  \right]
  \right\}
  \ ,
\\
\label{MaxwellZzphi}
  j^\pm_\phi
  &\equiv
  -
  {1 \over 8\pi \rho^3}
  {\vel \omegay \over \sqrt{\Deltay}}
  ( \Apot_x \pm \Apot_t )
  \left[
  {\partial \over \partial x}
  \ln
  \left(
  {1 \over \sigma^2} {\dd \omegax \over \dd x}
  \right)
  -
  z 
  \left(
  \pm
  {1 \over \sigma^2} {\dd \omegax \over \dd x}
  -
  {\vel \omegay \over \Deltay}
  \right)
  \right]
  \ ,
\end{align}
\end{subequations}
while the expression~(\ref{MaxwellZzeroy}) for $j^\pm_y$ remains unchanged.
The term proportional to $\Deltax$ on the far right hand side of
equation~(\ref{MaxwellZzxt})
is negligible compared to the other terms for $| \Deltax | \ll 1$,
so the radial Maxwell equation~(\ref{MaxwellZzxt}) simplifies to
\begin{equation}
\label{MaxwellZzxtsimp}
  j^\pm_x
  =
  \pm
  j^\pm_t
  =
  {1 \over 8\pi \rho^3}
  {\vel \over \sqrt{- \Deltax}}
  ( \Apot_x \pm \Apot_t )
  \left[
  \pm \,
  {\partial \over \partial x}
  \ln \left( {1 \over \sigma^2} {\dd \omegax \over \dd x} \right)
  -
  {\vel \omegay^2 \over \Deltay}
  \pm
  z
  {\vel \omegay \over \Deltay}
  \right]
  \ .
\end{equation}

\subsection{Collisionless source of electric current}

Maxwell's equations~(\ref{jk}) with currents given by
equations~(\ref{MaxwellZzxtsimp}), (\ref{MaxwellZzeroy}),
and (\ref{MaxwellZzphi})
can be satisfied by currents from
a sum of ingoing and outgoing collisionless streams,
equation~(\ref{jpm}),
with charge densities
\begin{equation}
\label{qNpmz}
  q^\pm N^\pm
  =
  -
  {\vel \Qelec^\pm \over 32\pi \rho^2 ( f_0 g_1 + f_1 g_0 )}
  {\dd \omegax \over \dd x}
  \left[
  {\partial \over \partial x}
  \ln \left( {1 \over \sigma^2} {\dd \omegax \over \dd x} \right)
  \mp
  {\vel \omegay^2 \over \Deltay}
  +
  z
  {\vel \omegay \over \Deltay}
  \right]
  \ ,
\end{equation}
and charge-weighted tetrad-frame momenta
\begin{equation}
\label{qppmz}
  p^\pm_k
  =
  {1 \over \rho}
  \left\{
  -
  {1 \over \sqrt{- \Deltax}}
  \, , \ 
  \mp
  {1 \over \sqrt{- \Deltax}}
  \, , \ 
  {1 \over \sqrt{\Deltay}}
  {
  \displaystyle
  {1 \over \sigma^2} {\dd \omegay \over \dd y}
  \over
  \displaystyle
  {\partial \over \partial x}
  \ln \left( {1 \over \sigma^2} {\dd \omegax \over \dd x} \right)
  \mp
  {\vel \omegay^2 \over \Deltay}
  +
  z
  {\vel \omegay \over \Deltay}
  }
  \, , \ 
  \pm
  {1 \over \sqrt{\Deltay}}
  {
  \displaystyle
  {\partial \over \partial x}
  \ln
  \left(
  {1 \over \sigma^2} {\dd \omegax \over \dd x}
  \right)
  -
  z 
  \left(
  \pm
  {1 \over \sigma^2} {\dd \omegax \over \dd x}
  +
  {\vel \omegay \over \Deltay}
  \right)
  \over
  \displaystyle
  {\partial \over \partial x}
  \ln \left( {1 \over \sigma^2} {\dd \omegax \over \dd x} \right)
  \mp
  {\vel \omegay^2 \over \Deltay}
  +
  z
  {\vel \omegay \over \Deltay}
  }
  \right\}
\ .
\end{equation}
The adjustable angle-dependent factor $z$ makes only a sub-dominant order $\vel$
change to the charge densities $q^\pm N^\pm$, equation~(\ref{qNpmz}),
but an order unity change to the azimuthal component $p^\pm_\phi$
of the momentum~(\ref{qppmz}).
Adjusting $z$ changes the azimuthal current $j^\pm_\phi$ arbitrarily,
but leaves the other current components essentially unchanged.

It is worth commenting that $z$ can be adjusted so that
the azimuthal current $j^\pm_\phi$ is zero for whichever stream is charged.
This in no way affects the conclusion that only one of the ingoing or outgoing
streams can be charged.
The divergence in the angular component~(\ref{Typhiem})
of the electromagnetic energy-momentum tensor
is driven by counter-streaming of radial, not angular,
ingoing and outgoing currents, coupled to the rotation of the black hole.
The only way to remove the divergence, as conformal separability requires,
is to allow only one of the radial currents to be charged.

\subsection{Energy-momenta}

A finite $Z_x \pm Z_t$ affects the radial electric field $\Er$,
equation~(\ref{EiBelec}),
to sub-dominant order $\vel$.
This small change propagates into corresponding components~(\ref{Tmnem})
of the electromagnetic energy-momentum tensor,
which leads to a small change in the
azimuthal component $p^\pm_\phi$ of the number-weighted tetrad-frame momenta
of collisionless ingoing and outgoing streams.
The small change amounts to changing
\begin{equation}
  {\partial \over \partial x}
  \ln
  \left(
  {1 \over \sigma^2} {\dd \omegax \over \dd x}
  \right)
  \rightarrow
  {\partial \over \partial x}
  \ln
  \left(
  {1 \over \sigma^2} {\dd \omegax \over \dd x}
  \right)
  +
  z {\vel \omegay \over \Deltay}
\end{equation}
in
expressions~(\ref{TxphiemU})
and (\ref{TxxmttemU})
for the electromagnetic energy-momentum components
$\Te_{x\phi} \pm \Te_{t\phi}$
and
$\Te_{xx} - \Te_{tt}$,
and in equation~(\ref{ppm}) for the azimuthal component $p^\pm_\phi$
of the number-weighted tetrad-frame momenta.

\end{document}